\documentclass[aps,prd,reprint,preprintnumbers,floats,epsfig,nofootinbib,amssymb,twocolumn,nofootinbib]{revtex4}

\def \lambdaeff{\lambda_\mathrm{eff}}
\usepackage{slashed}
\usepackage{graphicx,color}
\usepackage{epsfig}
\usepackage{subfigure}
\usepackage{epsfig}
\usepackage{dcolumn}
\usepackage{bm}
\usepackage{color}
\usepackage{ulem}
\usepackage{enumitem}
\begin{document}

\preprint{ACFI-T18-17}


\title{Scalar Electroweak Multiplet  Dark Matter}

\author{Wei Chao$^1$}
\email{chaowei@bnu.edu.cn}
\author{Gui-Jun Ding$^2$}
\email{dinggj@ustc.edu.cn}
\author{Xiao-Gang He$^{3,4,5}$}
\email{hexg@phys.ntu.edu.tw}
\author{Michael Ramsey-Musolf$^{6,7}$}
\email{mjrm@physics.umass.edu}
\address{$^1$Center for advanced quantum studies, Department of Physics, Beijing Normal University, 100875, Beijing, China \\
$^2$ Interdisciplinary Center for Theoretical Study and Department of Modern Physics,
University of Science and Technology of China, Hefei, Anhui 230026, China  \\
$^3$Tsung-Dao Lee Institute, and School of Physics and Astronomy, Shanghai Jiao Tong University, Shanghai 200240, China\\
$^4$Department of Physics, National Taiwan University, Taipei 106, Taiwan \\
$^5$National Center for Theoretical Sciences, Hsinchu 300, Taiwan \\
$^6$ Amherst Center for Fundamental Interactions, University of Massachusetts-Amherst, Department of Physics, Amherst, MA 01003, USA\\
$^7$ Kellogg Radiation Laboratory, California Institute of Technology, Pasadena, CA 91125 USA}

\begin{abstract}

We revisit the theory and phenomenology of scalar electroweak multiplet thermal dark matter. We derive the most general, renormalizable scalar potential, assuming the presence of the Standard Model Higgs doublet, $H$, and an electroweak multiplet $\Phi$ of arbitrary SU(2$)_L$ rank and hypercharge, $Y$. We show that, in general, the $\Phi$-$H$ Higgs portal interactions depend on three, rather than two independent couplings as has been previously considered in the literature. For the phenomenologically viable case of $Y=0$ multiplets, we focus on the septuplet and quintuplet cases, and consider the interplay of relic density and spin-independent direct detection cross section. We show that both the relic density and direct detection cross sections depend on a single linear combination of Higgs portal couplings, $\lambda_{\rm eff}$. For $\lambda_{\rm eff}\sim \mathcal{O}(1)$, present direct detection exclusion limits imply that the neutral component of a scalar electroweak multiplet would comprise a subdominant fraction of the observed DM relic density.

\end{abstract}

\maketitle

\section{Introduction}
Determining the identity of the dark matter and the nature of its interactions is a forefront challenge for astroparticle physics. A plethora of scenarios have been proposed over the years, and it remains to be seen whether any of these ideas is realized in nature. One hopes that results from ongoing and future dark matter direct and indirect detection experiments, in tandem with searches for dark matter signatures at the Large Hadron Collider and possible future colliders, will eventually reveal the identity of dark matter and the character of its interactions.

A widely studied possibility of continuing interest is that dark matter consists of weakly interacting massive particles (WIMPs). An array of realizations of the WIMP paradigm have been considered, ranging from ultraviolet complete theories such as the Minimal Supersymmetric Standard Model to simplified models containing a relatively small number of degrees of freedom and new interactions. In the latter context, one may classify WIMP dark matter candidates according to their spin and electroweak gauge quantum numbers. The simplest possibility involves SU(2$)_L\times$U(1)$_Y$ gauge singlets. Null results from direct detection (DD) experiments and LHC searches place severe constraints on this possibility, though some room remains depending on the specific model realization.

An alternative possibility is that the dark matter consists of the neutral component of an electroweak multiplet, $\chi^0$. A classification of these possibilities is given in\cite{Cirelli:2005uq}. Those favored by the absence of DD signals carry zero hypercharge ($Y$), thereby preventing overly-large WIMP-nucleus cross sections mediated by $Z^0$ exchange. Stability of the $\chi^0$ requires imposition of a discrete ${\bm Z}_2$ symmetry unless the representation of the electroweak multiplet is of sufficiently high dimension: d=5 for fermions and d=7 for scalars. These scenarios with sufficiently high dimension representation go under the heading \lq\lq minimal dark matter".

In this work, we consider features of scalar electroweak multiplet dark matter $\Phi$, including but not restricting our attention to minimal dark matter (as conventionally defined). The phenomenology of scalar triplet dark matter, involving a multiplet transforming as $(1, 3, 0)$ under SU(3$)_C$ and electroweak symmetries, has been considered previously in Refs.~\cite{FileviezPerez:2008bj,Lu:2016dbc,Fischer:2013hwa,JosseMichaux:2012wj,Basak:2012bd,Araki:2011hm,Pal:1987mr}. Extensive studies for  other electroweak multiplets of dimension $n$ have been reported in Refs.~\cite{Hambye:2009pw,AbdusSalam:2013eya}. The authors of Ref.~\cite{Hambye:2009pw} considered the Inert Doublet model and the $n=3,5,7$ scalar electroweak multiplets and discussed the impact of non-vanishing Higgs portal interactions on the relic density, spin-independent dark matter-nucleus cross section, $\sigma_{SI}$, and indirect detection (ID) signals. Ref.~\cite{AbdusSalam:2013eya} also considered the impact of Higgs portal interactions on the relic density and $\sigma_{SI}$ but did not analyze the implications for indirect detection. The latter study also focused on a relatively light mass for the dark matter candidate, for which it would appear to undersaturate the relic density.

In what follows, we revisit the topic of these earlier studies, taking into account several new features that may require modifying some of the conclusions in Refs.~\cite{Hambye:2009pw,AbdusSalam:2013eya}:

\begin{itemize}
\item We find that the scalar potentials $V(H,\Phi)$ given in Refs.~\cite{Hambye:2009pw,AbdusSalam:2013eya}  are not the most general renormalizable potentials and that, depending on the representation of SU(2$)_L\times$U(1$)_Y$ there exist one or more additional interactions that should be included. For the $Y=0$ representations, the  $\Phi$-$H$ interaction relevant for both the relic density and DD cross section involves an effective coupling $\lambda_\mathrm{eff}$ that is linear combination of two of the three possible Higgs portal couplings. The specific linear combination is representation dependent.
\item We update the computation of $\sigma_{SI}$ taking into account the nucleon matrix elements of twist-two operators generated by gauge boson-mediated box graph contributions as outlined in Refs.~\cite{Hisano:2011cs,Hisano:2014kua,Hill:2014yka,Hill:2014yxa}. We note that Ref.~\cite{AbdusSalam:2013eya} considered only the Higgs portal contribution to $\sigma_{SI}$ and did not include the effect of electroweak gauge bosons. We find that the gauge boson-mediated box graph contributions are smaller in magnitude that given in Ref.~\cite{Hambye:2009pw}, which used the expressions given in Ref.~\cite{Cirelli:2005uq}. In general, the Higgs portal contribution dominates the DD detection cross section except for very small values of $\lambda_\mathrm{eff}$.
\item The presence of a non-vanishing $\lambdaeff$ can allow for a larger maximum dark matter mass, $M$, to be consistent with the observed relic density than one would infer when considering only gauge interactions. For the cases we consider below, this maximum mass be as larger as $\mathcal{O}(20)$ TeV for perturbative values of $\lambdaeff$.
\item For moderate values of the Higgs portal couplings, the spin-independent cross section, scaled by the fraction of the relic density comprised by $\Phi^0$, is a function $\lambdaeff$ and  $M$. The present DD bounds on $\sigma_\mathrm{SI}$ generally require $M\lesssim 5$ TeV for perturbative values of $\lambdaeff$ -- well below the maximum mass consistent with the observed relic density.
\end{itemize}

In what follows, we provide the detailed analysis leading to these conclusions. For the structure of $V(H,\Phi)$ we consider $\Phi$ to be a general representation of SU(2$)_L\times$U(1$)_Y$. Previous studies have considered in detail electroweak singlets ($n=1$), doublets ($n=2$), and triplets ($n=3$). In all three cases, stability of the DM particle requires that one impose a discrete symmetry on the Lagrangian. Going to higher dimension representations, it has been shown in Ref.~\cite{Cirelli:2005uq} that for $n=4$, stability of the neutral component also requires imposition of a discrete symmetry, while for $n=5$, the neutral component can only decay through a non-renormalizable dimension five operator with coefficient suppressed by one power of a heavy mass scale $\Lambda$. In the latter case, it is possible to ensure DM stability  on cosmological time scales by either imposing a discrete symmetry or by choosing $\Lambda$ to be well above the Planck scale. At $n=7$, the first non-renormalizable, decay-inducing operator appears at higher dimension, and DM stability may be ensured even without imposition of a discrete symmetry by choosing $\Lambda$ below the Planck scale.
\\








With the foregoing considerations in mind, we focus on the $n=5$ and $7$ cases for purposes of illustrating the dark matter phenomenology.
Since the group theory relevant to construction of $V(H,\Phi)$ is rather involved, we provide a detailed discussion in Appendices \ref{sec:groupA} and \ref{sec:groupB}. In Section \ref{sec:models}, we start with a general formulation, followed by treatment of specific model cases. Section \ref{sec:relic} gives the calculation of the relic density, including the effects of coannihilation and the Sommerfeld enhancement. We compute $\sigma_{SI}$ in Section \ref{sec:direct}. We summarize in Section \ref{sec:conclude}. Along the way, we point out where we find differences with earlier studies.

\section{Models}
\label{sec:models}

We consider the renormalizable Higgs portal interactions involving $H$ and $\Phi$ for two illustrative cases. We restrict our attention to $\Phi$ being a complex scalar with $Y=0$. The form of the potential for $\Phi$ being a real representation of SU(2$)_L$ with $Y=0$ is relatively simple. The corresponding features have been illustrated in previous studies wherein $\Phi$ is either an SU(2$)_L$ singlet or real triplet. Consequently, we focus on complex representations, using the $n=5$ and $n=7$ examples, to illustrate the new features not considered in earlier work.

To proceed, we first introduce some notation. It is convenient to consider both $\Phi$ and the associated conjugate ${\bar\Phi}$, whose components are related to those of $\Phi$ as
\begin{equation}
{\overline\Phi}_{j,m} = (-1)^{j-m} \Phi^\ast_{j,-m}\ \ \ .
\end{equation}
As we discuss in Appendix \ref{sec:groupA}, $\Phi$ and ${\overline\Phi}$ transform in the same way under SU(2$)_L$. One may then proceed to build SU(2$)_L$ invariants by first coupling $\Phi$, $\overline{\Phi}$, $H$, and $\overline{H}$ pairwise into irreducible representations and finally into SU(2$)_L$ invariants. For example,
\begin{equation}
\left(\Phi{\overline\Phi}\right)_0 = \frac{(-1)^{2j}}{\sqrt{2j+1}}\, \Phi^\dag\Phi\ \ \ ,
\end{equation}
which in general is a distinct invariant from $\left(\Phi{\Phi}\right)_0$ except in special cases when $\Phi$ is a real scalar multiplet satisfying $\Phi=\overline{\Phi}$. Note that for $j=1/2$ , $\left(\Phi{\Phi}\right)_0$ vanishes, so that there is only one quadratic invariant in this case as well. Quartic interactions can be constructed in a variety of ways, such as
\begin{equation}
\left[\left(\Phi{\Phi}\right)_J \left({\overline\Phi}\,{\overline\Phi}\right)_J\right]_0
\end{equation}
for $\Phi$ self-interactions or
\begin{equation}
\left[\left({\overline H} H\right)_L \left({\overline \Phi} \Phi\right)_L \right]_0
\label{eq:portal1}
\end{equation}
with $L={0,1}$ for the Higgs portal interactions. Note that there exists a third such interaction
\begin{equation}
\left({\overline H} H\right)_0 \left({\Phi} \Phi\right)_0
\label{eq:portal2}
\end{equation}
that is distinct from the $L=0$ operator in Eq.~(\ref{eq:portal1}) for $\Phi$ being a complex integer representation. We note that previous studies have not in generally included all three of the possible Higgs portal interactions. The classification of the $\Phi$ self-interactions is more involved, and it is most illuminating to consider them on a case-by-case basis.

\subsection{Setptuplet}

The interactions can be written as
\begin{eqnarray}
V&=& + M_A^2 (\Phi^\dagger \Phi ) + \left\{ M_B^2 ( \Phi \Phi)_0^{} + {\rm h.c.} \right\} -\mu^2 H^\dagger H  \nonumber
\\ && + \lambda ( H^\dagger H)^2 + \lambda_1^{} ( H^\dagger H ) (\Phi^\dagger \Phi )
\\ && + \lambda_2^{} [(\overline H H)_1^{} (\overline \Phi \Phi )_1^{} ]_0^{}   +   [{\lambda_3}(\overline H H)_0^{} (\Phi \Phi)_0^{} + {\rm h.c. }] \; , \nonumber
\end{eqnarray}
where $H$ is the Higgs doublet and $\Phi$ is a complex electroweak septuplet with
\begin{eqnarray}
(\Phi \Phi)_0^{} &=& {1\over \sqrt{7}}\sum_{m=-3}^3    (-1)^{{3-}m} \phi_{3,m}^{}  \phi_{3,-m}^{} \nonumber
\\&=& {1\over \sqrt{7}} \bigl(2 \phi_{3,3} \phi_{3,-3}{-}2 \phi_{3,2} \phi_{3,-2} \nonumber
\\ && {+}2 \Phi_{3,1} \phi_{3,-1}
{-}\phi_{3,0} \Phi_{3,0}  \bigr) \\
(\overline{H} H)_0^{} &=&{1\over \sqrt{2}}\left[ (H^+)^* H^+ + (H^0)^* H^0 \right]
\end{eqnarray}
and
\begin{eqnarray}
(\overline H H)_1^{} &=& \left( \matrix{ ( H^0)^* H^+ \cr  {1\over \sqrt{2} } \left[( H^0)^*H^0 - (H^+)^* H^+ \right] \cr -(H^+)^* H^0 } \right) \\
(\overline \Phi \Phi )_1^{} &=& \left( \matrix{  {1 \over {14} } \mathcal{A} \cr -{\sqrt{7} \over 14 } \sum_{m=-3}^3 m \phi_{3, m}^* \phi_{3, m}   \cr {1 \over 14}\mathcal{B} }\right)
\end{eqnarray}
with
\begin{eqnarray}
\mathcal{A} &=&+\sqrt{21} \phi_{3, -3}^* \phi_{3, -2}^{} + \sqrt{35} \phi_{3,-2}^* \phi_{3,-1}^{}  + \sqrt{42} \phi_{3,-1}^* \phi_{3,0}^{}\nonumber \\&&+ \sqrt{42} \phi_{3,0}^* \phi_{3,1}^{}  +\sqrt{35} \phi_{3,1}^* \phi_{3, 2}^{} + \sqrt{21} \phi_{3,2}^* \phi_{3,3}^{}  \\
\mathcal{B}&= & -\sqrt{21} \phi_{3,-2}^* \phi_{3,-3}^{} -\sqrt{35} \phi_{3,-1}^* \phi_{3, -2}^{}  -\sqrt{42} \phi_{3,0}^* \phi_{3,-1}^{} \nonumber \\ &&- \sqrt{42} \phi_{3, 1}^* \phi_{3, 0}^{} -\sqrt{35} \phi_{3,2}^* \phi_{3, 1}^{} -\sqrt{21} \phi_{3,3}^* \phi_{3, 2}^{}
\end{eqnarray}
After electroweak symmetry breaking, wherein
\begin{equation}
\mathrm{Re}\, H^0\rightarrow \left(v+h\right)/\sqrt{2}
\end{equation}
one obtains the $\Phi$ mass term
\begin{widetext}
\begin{eqnarray}
\mathcal{L}_\mathrm{mass} = \left(  \matrix { \phi_{3, k }^{}  & \phi_{3, -k}^* }\right) \left( \matrix{M_A^2 + {1\over 2 }\lambda_1^{} v^2+{1 \over 4\sqrt{42}} k \lambda_2 v^2  &  {\sqrt{7} \over  7} (-1)^{k+1}   \left \{ 2M_B^2 + {1 \over \sqrt{2}}   \lambda_3^{} v^2 \right\}\cr  {\sqrt{7} \over  7} (-1)^{k+1}   \left\{ 2M_B^{2*} + {1\over \sqrt{2}}  \lambda_3^{*} v^2 \right\}&M_A^2 + {1\over 2 }\lambda_1^{} v^2-{1 \over 4\sqrt{42}} k \lambda_2 v^2 } \right) \left(  \matrix{ \phi_{3, k}^* \cr \phi_{3,-k}^{} }\right)
\end{eqnarray}
\end{widetext}
By setting $\phi_{3, 0} =(\phi_{3; (0, +)}+ i \phi_{3; (0,-)})/\sqrt{2} $, the neutral scalar mass matrix can be written as
\begin{widetext}
\begin{eqnarray}
\label{eq:massSQ_neutral}\left( \matrix{  M_A^2 +{1\over 2} \lambda_1^{} v^2  -{ 2\over \sqrt{7}} {\rm Re} (M_B^2 )  - {1 \over \sqrt{14}} {\rm Re} (\lambda_3 ^{} ) v^2 & {2 \over \sqrt{7}} {\rm Im } (M_B^2)+ {1\over \sqrt{14}} {\rm Im } (\lambda_3^{} ) v^2  \cr  {2 \over \sqrt{7}} {\rm Im } (M_B^2)+ {1\over \sqrt{14}} {\rm Im } (\lambda_3^{} ) v^2  &    M_A^2+{1\over 2} \lambda_1^{} v^2   +{ 2\over \sqrt{7}} {\rm Re} (M_B^2 )  +{1 \over \sqrt{14}} {\rm Re} (\lambda_3 ^{} ) v^2 } \right)
\end{eqnarray}
\end{widetext}
in the basis $(\phi_{3; (0,+)},~\phi_{3; (0,-)})^T$.
{Then we have the mass eigenvalues}
\begin{widetext}
\begin{eqnarray}
M_{ {\hat\phi}_{3; \pm k}}^2 &=& M_A^2 + {1\over 2 } \lambda_1^{} v^2 \pm \sqrt{ \left|{2 \over \sqrt{7}} M_B^2 + {1\over \sqrt{14}} \lambda_3^{} v^2 \right|^2 + \left( {1\over 4\sqrt{42} } k \lambda_2^{} v^2 \right)^2  }
\label{eq:massA} \\
M_{{\hat\phi}_{3; (0,\pm)}}^2 &=&  M_A^2 + {1\over 2 } \lambda_1^{} v^2 \pm  \left|{2 \over \sqrt{7}} M_B^2 + {1\over \sqrt{14}} \lambda_3^{} v^2 \right|
\label{eq:massB}
\end{eqnarray}
\end{widetext}
{ where for each isospin projection $k$, the \lq\lq $\pm$ denotes the upper or lower sign in Eqs.~(\ref{eq:massA},\ref{eq:massB}) and where the notation ${\hat\phi}_{3,\pm k}$ indicates the mass eigenstate. }

{
{From these expressions we conclude that }
\begin{itemize}
\item { If $\lambda_2$ is nonzero,
there will be no dark matter since one may have $M_{ \phi_{3; (k, -)}}^2 < M_{ \phi_{3; (0, -)}}^2 $ for $k\not= 0$. One needs $\lambda_2 \sim 0$, otherwise there may  exist long-lived charged scalars. }
\item { For $\lambda_2 =0$,  we have two real septuplets
\begin{eqnarray}
 S_A^{}={1\over \sqrt{2}}\left( \matrix{ \hat\phi_{3,-3}^* \cr i{\hat\phi}_{3,-2}^* \cr \hat\phi_{3,-1}^* \cr i {\hat\phi}_{3; (0,-)}^{} \cr \hat\phi_{3,-1} \cr i{\hat\Phi}_{3,-2} \cr \hat\phi_{3,-3}    }\right)   \hspace{0.5cm}
  S_B^{}={1\over\sqrt{2}}\left( \matrix{ {\hat\phi}_{3,3} \cr i\hat\phi_{3,2} \cr {\hat\phi}_{3,1} \cr i {\hat\phi}_{3; (0,+)}^{} \cr {\hat\phi}_{3,1}^* \cr i\hat\phi_{3,2}^* \cr {\hat\phi}_{3,3}^*    }\right)
\end{eqnarray}
The corresponding mass eigenvalues eigenvalues are
\begin{eqnarray}
M_{S_A^{}, S_B^{}}^2 &=&  M_A^2 + {1\over 2 } \lambda_1^{} v^2 \\
&& \pm \left|{2 \over \sqrt{7}} M_B^2 + {1\over \sqrt{14}} \lambda_3^{} v^2 \right| \; ,
\nonumber
\end{eqnarray}
where the lower (upper) sign corresponds to $S_A$ ($S_B$)}.
\item In general, the neutral component of $S_A$ -- denoted here as the real scalar $\chi$ -- will be the DM particle. Radiative corrections will give rise to the mass splitting between the neutral and charged components. In the limit $M_A^{} \gg M_{W,Z}^{}$, one has $M_Q-M_0\approx Q^2 \Delta M$, with $\Delta M =(166\pm1 )~{\rm MeV}$~\cite{Cirelli:2005uq} being the mass splitting between the Q=1 and 0 components.

\end{itemize}
%
%

 From the full scalar potential, one may obtain dark matter self interactions
 \begin{equation}
 \mathcal{L}_\chi^\mathrm{self} = -{\tilde \lambda}_\mathrm{self}\,  \chi^4 \ \ \ ,
 \end{equation}
 which may be important in solving the core-cusp problem~\cite{deBlok:2009sp,Tulin:2017ara}. The relevant terms are
\begin{eqnarray}
&&\sum_{J=0}^{2J} \kappa_k^{} [(\Phi \Phi)_k (\overline{\Phi}\,\overline{\Phi})_k]_0^{} +\sum_{k=0}^{2J} \left\{ \kappa_k^{\prime} [(\Phi \Phi)_k^{}  (\Phi  \Phi)_k^{}]_0^{} \right. \nonumber \\ && \left.+\kappa_k^{\prime\prime} [(\overline{\Phi}\, \Phi)_k^{} (\Phi  \Phi)_k^{}]_0^{} + {\rm h.c.} \right\}
\label{eq:selfquartic}
\end{eqnarray}
Note that each component of $\left(\Phi\Phi\right)_{j}$ $(j=0,\ldots,6)$ is determined by
\begin{eqnarray}
\left(\Phi\Phi\right)_{j,m}=\sum_{m_1,m_2}C^{j,m}_{3,m_1;3,m_2}\phi_{3,m_1}\phi_{3,m_2}\,.
\end{eqnarray}
From the property of Clebsch-Gordan coefficients:
\begin{equation}
C^{j,m}_{j_1,m_1;j_2,m_2}=(-1)^{j-j_1-j_2}C^{j,m}_{j_2,m_2;j_1,m_1}\,.
\end{equation}
If $j-j_1-j_2$ is an odd (even) integer, the corresponding contraction of two $\Phi$ fields is antisymmetric (symmetric). Consequently, $\left(\Phi\Phi\right)_{1}$,  $\left(\Phi\Phi\right)_{3}$ and $\left(\Phi\Phi\right)_{5}$ vanish.
For the most general case leading to the mass-squared matrix in Eq.~(\ref{eq:massSQ_neutral}), the expression for the DM quartic self interaction is rather involved and not particularly enlightening. For completeness, in Appendix \ref{sec:groupC} we give an expression for the quartic interactions in terms of $\phi_{3;(0,\pm)}$, from which one can determine the DM self interaction by expressing the $\phi_{3;(0,\pm)}$ in terms of the mass eigenstates. To illustrate, we give here the result for the special case of real $M_B^2$ and $\lambda_3$ with $2\sqrt{2}M^2_B + \lambda_3 v^2<0$:
\begin{eqnarray}
4\tilde \lambda_{\rm self}^{} &=& +{1\over 7} \left[\kappa_0+ 2 {\rm Re} (\kappa_0^\prime)+2{\rm Re} (\kappa_0^{\prime \prime})\right] \nonumber \\ &&+{4\over 21\sqrt{5}} \left[\kappa_2+ 2 {\rm Re} (\kappa_2^\prime)+2{\rm Re} (\kappa_2^{\prime \prime})\right] \nonumber\\
&&+{6\over 77} \left[\kappa_4+ 2 {\rm Re} (\kappa_4^\prime)+2{\rm Re} (\kappa_4^{\prime \prime})\right] \nonumber \\&&+{100 \over 231 \sqrt{13}}  \left[\kappa_6+ 2 {\rm Re} (\kappa_6^\prime)+2{\rm Re} (\kappa_6^{\prime \prime})\right]\,,
\end{eqnarray}
where the factor $4$ comes from the fact that {$\phi_{3, 0} =(\phi_{3; (0, +)}+ i \phi_{3; (0,-)})/\sqrt{2}$}.   In general, ${\tilde\lambda}_\mathrm{self}$ depends on 12 free parameters in Eq.~(\ref{eq:selfquartic}).    We defer an exploration of the possible additional physical consequences of these independent interactions to future work.

}

{

\subsection{Quintuplet}
The analysis for the electroweak scalar quintuplet dark matter is similar to the septuplet case. For purposes of completeness, we include some of the important features below.  The complex quintuplet scalar field with $j=2$ and $Y=0$ is denoted by
\begin{equation}
\Phi^{}=\left(\begin{array}{c}
\phi_{2,2}\\
\phi_{2,1}\\
\phi_{2,0}\\
\phi_{2,-1}\\
\phi_{2,-2}\\
\end{array}
\right)\,.
\end{equation}
The mass term and interactions of quintuplet are the same as those of the septuplet given in Eq. (6), where we set  $\lambda_2=0$ to ensure the presence of  a stable neutral component.
To derive the mass eigenvalues we consider the contractions of the two scalar multiplets $\Phi\Phi$. According to general decomposition rule, one has
\begin{eqnarray}
&&\left(\Phi^{}\Phi^{}\right)_{0}=\frac{\phi_{2,0}^2-
   2 \phi_{2,-1}
   \phi_{2,1}+2
   \phi_{2,-2}
   \phi_{2,2}}{\sqrt{5}}\; .
\end{eqnarray}
By setting $\phi_{2,0}= (\alpha^\prime+ i \beta^\prime)/\sqrt{2}$, the mass matrix of the neutral scalars can be written as
\begin{widetext}
\begin{eqnarray}
{1\over 2 }\left( \matrix{\alpha^\prime & \beta^\prime}\right) \left( \matrix{ M_A^2 +{1\over 2} \lambda_1^{} v^2  + {2 \over \sqrt{5} } {\rm Re }M_B^2 -{1\over \sqrt{10} }{\rm Re} (\lambda_3^{} ) v^2  & -{2\over \sqrt{5}} {\rm Im} (M_B^2) +{1\over \sqrt{10} } {\rm Im}(\lambda_3^{} )v^2\cr -{2\over \sqrt{5}} {\rm Im} (M_B^2) +{1\over \sqrt{10} } {\rm Im}(\lambda_3^{} )v^2& M_A^2 +{1\over 2} \lambda_1^{} v^2-  {2 \over \sqrt{5} } {\rm Re }M_B^2 + {1\over \sqrt{10} }{\rm Re} (\lambda_3^{} ) v^2 } \right) \left(\matrix{\alpha^\prime \cr \beta^\prime} \right)
\end{eqnarray}
\end{widetext}
The mass eigenvalues are
\begin{eqnarray}
M^2_{\alpha^\prime, \beta^\prime} = M_A^2 +{1\over 2 } \lambda_1^{} v^2 \pm \left|{2\over \sqrt{5}} M_B^2 - {1\over \sqrt{10}} \lambda_3^{} v^2 \right| \; ,
\end{eqnarray}
which are also mass eigenvalues of the two real quintuplet.

The self-coupling can be derived following the same strategy of the septuplet case, and we give the results in  Appendix \ref{sec:groupC}.

}

\section{Relic density}
\label{sec:relic}

In this work, we assume that dark matter in the early Universe was in the local thermodynamic equilibrium. Decoupling ocurred when its interaction rate drops below the expansion rate of the Universe.  The corresponding evolution of the dark matter number density $n$, is governed by the Boltzmann equation:
\begin{eqnarray}
\dot{n} + 3 Hn =- \langle \sigma v_{\rm M\slashed{o}ller} \rangle ( n^2 -n_{\rm EQ}^2 ) \; , \label{boltzmann}
\end{eqnarray}
where $H $ is the Hubble constant, $\sigma v_{\rm M\slashed{o}ller}$ is the total annihilation cross section multiplied by  the M$\slashed{\rm o}$ller velocity, $v_{\rm M\slashed{o}ller}=(|v_1 -v_2 |^2 -|v_1 \times v_2 |^2 )^{1/2}$, brackets denote thermal average and $n_{\rm EQ}$ is the number density at thermal equilibrium. It has been shown that
\begin{equation}
\langle \sigma v_{\rm M\slashed{o}ller}\rangle  =\langle \sigma v_{\rm lab} \rangle = \frac{1}{2} [1 + K_1^2 (x) /K_2^2 (x)] \langle \sigma v_{\rm cm} \rangle\  ,
\end{equation}
where $x=m/T$, $K_i$ are the modified Bessel functions of order $i$.

In a general framework that includes co-annihilation, the dynamics depend on a set of species $\{\chi_i\}$ with masses $\{m_i\}$ and number densities $\{n_i\}$.
It has been shown that the total number density of all species taking part in the co-annihilation process, $n\equiv \sum_i n_i$,  obeys Eq. (\ref{boltzmann}).
In this case $\langle\sigma v_{\rm M\slashed{o}ller}\rangle$ can be written as~\cite{Edsjo:1997bg,Nihei:2002sc}
\begin{eqnarray}
\langle \sigma v_{\rm M\slashed{o}ller} \rangle = {\int_{4m_\chi^2}^{\infty} ds s^{3/2} K_1 \left( {\sqrt{s}\over T} \right) \sum_{ij}^N \beta_{ij}^2 {g_i g_j \over g_\chi^2 } \sigma_{ij} (s)
\over
8 m_\chi^4 T\left[ \sum_i^N {g_i^{} \over g_\chi^{}} {m_i^2 \over m_\chi^2 } K_2^2 \left( {m_i \over T}\right)\right]^2} \label{co-annihilation}
\end{eqnarray}
where $g_i$ is the number of degrees of freedom, $s$ is the Mandelstam variable, $\sigma_{ij} =\sigma(\chi_i \chi_j \to \text{all})$, and the kinematic factor $\beta_f (s, m_i, m_j)$ is given by
\begin{eqnarray}
\beta_{ij} = \sqrt{  \left[1- {(m_i + m_j)^2 \over s} \right]\left[1- {(m_i - m_j)^2 \over s}\right]} \; .
\end{eqnarray}
The number density of the dark matter at the end will be $n_\chi=n$.
%
The relic density of the dark matter today can be written as
\begin{eqnarray}
\Omega_\chi h^2 &=& {1.66 T_\gamma^3 \sqrt{g_*} \over M_{\rm pl} \rho_{\rm crit}} \left( {T_\chi\over T_\gamma}\right)^3    \left[\int_0^{x_f} dx \langle \sigma v_{\rm M\slashed{o}ller} \rangle (x) \right]^{-1}
\end{eqnarray}
where $\rho_{\rm crit}\equiv 1.05\times10^{-5}(h^2) ~{\rm GeV}/cm^3$ is the critical density, $M_{\rm pl} $ denotes the Planck mass, $T_\gamma$ and $T_\chi$ are the present temperatures of  photon and dark matter, respectively. According to entropy conservation in a comoving volume,  the suppression factor  $(T_\chi/T_\gamma)^3\approx 1/20$\cite{Olive:1980wz}.

\subsection{The  single species case}

We first calculate the dark matter relic density assuming only a single species, {\it i.e.}, including no co-annihilation. To show the interplay between the Higgs portal and gauge interactions in the annihilation dynamics, we compute the relic density analytically. For completeness, we show the
thermal average of various annihilation cross sections:
\begin{widetext}
\begin{eqnarray}
\langle\sigma v \rangle_{ hh}^{}  &= &\lambda_{\rm eff}^2  \left\{ {\sqrt{m^2 -m_h^2 }\over 32 \pi m^3} + {-4m^2 + 5 m_h^2  \over 256 \pi  m^3 \sqrt{m^2 -m_h^2 }} \langle  v^2\rangle  \right \} \label{a} \\
\langle\sigma v \rangle_{ \bar tt }^{}  &= &\lambda_{\rm eff}^2 \left\{   {  m_t^2 (m^2 -m_t^2)^{3/2}  \over 4 \pi m^3 (4 m^2 -m_h^2)^2 } + \Delta_1 (t) \langle v^2 \rangle \right\}  \label{b}\\
\langle\sigma v \rangle_{ ZZ}^{}  &= &\lambda_{\rm eff}^2  \left\{ { \sqrt{m^2 -m_z^2 } (4 m^4 - 4 m^2 m_z^2 +3 m_z^4 ) \over 8 \pi m^3 (4 m^2 -m_h^2 )^2 }  + \Delta_2 (Z) \langle  v^2 \rangle   \right\} \label{c}\\
\langle\sigma v \rangle_{ WW }^{}  &= &\lambda_{\rm eff}^2  \left\{ { \sqrt{m^2 -m_w^2 } (4 m^4 - 4 m^2 m_w^2 +3 m_w^4 ) \over 8 \pi m^3 (4 m^2 -m_h^2 )^2 }  + \Delta_2 (W) \langle  v^2 \rangle   \right\}  \nonumber \\
&+&\lambda_{\rm eff}^{} {9 g^2 v^2  \over  16 \pi m^2( 4m^2 -m_h^2 ) }  + {c_{n}^2 g^4\over 4 \pi m^2 } \label{d}
\end{eqnarray}
\end{widetext}
where $\lambda_\mathrm{eff}$ is an effective coupling given by a linear combination of the independent Higgs portal couplings. Assuming real $M_B^2$ and $\lambda_3$ one has

 \begin{eqnarray}
\lambda_\mathrm{eff} = \cases { \lambda_1\pm \sqrt{\frac{2}{7}} \lambda_3\ , & {septuplet}\cr
\lambda_1\mp\sqrt{\frac{2}{5}} \lambda_3\ ,&  {quintuplet}}\,,
\label{eq:lambdaeff}
\end{eqnarray}
where we have set $\lambda_2=0$ as above; where the upper (lower) signs correspond to $2\sqrt{2} M_B^2+\lambda_3 v^2$ being negative (positive);
where the parameter 
\begin{equation}
c_n = \frac{(n^2-1)^2}{64}
\end{equation}
accounts for the effective couplings of the dark matter with the $W$ boson; and where
\begin{widetext}
\begin{eqnarray}
\Delta_1 (t)& =&{ m_t^2 \sqrt{m^2-m_t^2 } (-24 m^4 -5 m^2 m_t^2  + 2 m^2 (m_h^2 + 18 m_t^2 )) \over 32 \pi  m^3 (4m^2 -m_h^2)^3} \\
\Delta_2(v) &=& { -64 m^8 +176 m^6 m_v^2 -15 m_h^2 m_v^6 - 4m^4 (3 m_h^2 m_v^2 + 52 m_v^4) + 12 m^2 (2 m_h^2 m_v^4 + 9 m_v^6) \over 64 \pi m^3 ( 4 m^2 -m_h^2)^3 \sqrt{m^2 - m_v^2}}
\end{eqnarray}
\end{widetext}

The present relic density of the DM is simply given by $\rho_\chi = M  n_\chi$.
The relic density can finally be expressed in terms of the critical density
\begin{eqnarray}
\Omega h^2 \approx {1.07 \times 10^9 {\rm GeV}^{-1}  x_F \over M_{pl} \sqrt{g_*} ( a + 3 b/x_F )} \; ,
\end{eqnarray}
\begin{figure}[t]
\begin{center}
\includegraphics[width=7.5cm,height=7cm,angle=0]{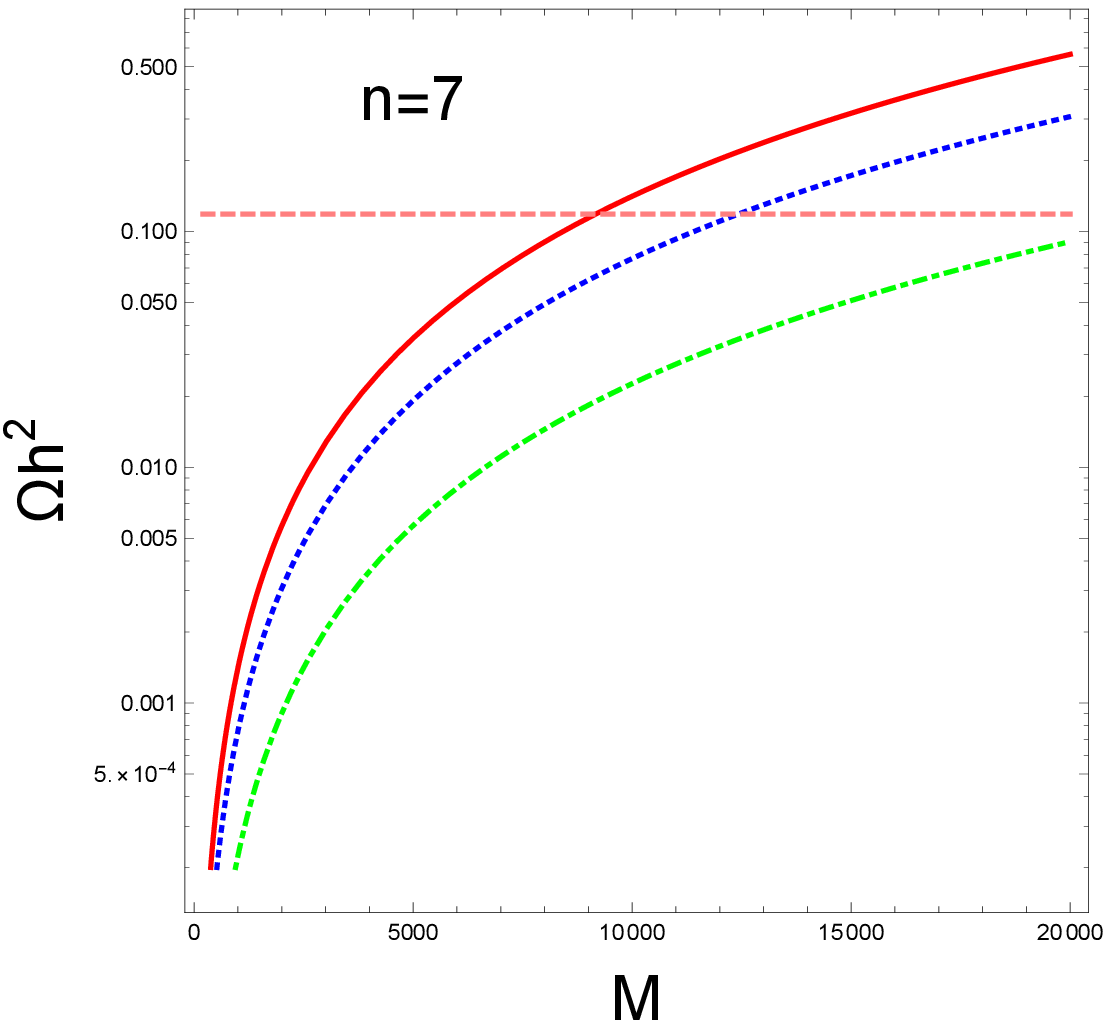}
\includegraphics[width=7.5cm,height=7cm,angle=0]{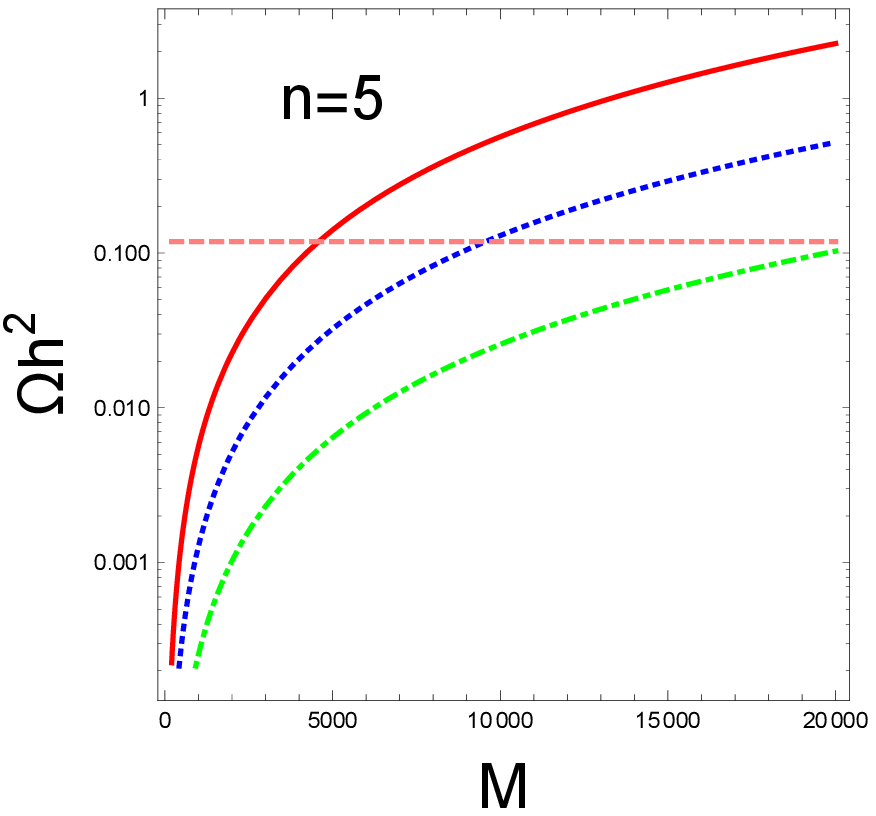}
\end{center}
\caption{Dark matter relic density as a function of the dark matter mass.  The solid (red),  dashed (blue), and dot-dashed (green) curves correspond to $\lambda_\mathrm{eff} =0,2,5$, respectively.  The horizontal line is the observed relic density.  }  \label{density}\end{figure}
where  $a$ and $b$, which are given in  Eqs. (\ref{a}-\ref{d}), are  expressed in ${\rm GeV}^{-2}$ and $g_*$ is the effective degrees of freedom at the freeze-out temperature $T_F$, $x_F= M/T_F $, which can be estimated through the iterative solution of the equation
\begin{eqnarray}
x_F = \ln \left[ c(c+2) \sqrt{45 \over 8 } {g\over 2 \pi^3 } { M M_{pl} (a + 6 b/x_F ) \over \sqrt{g_* x_F }}\right] \; ,
\end{eqnarray}
where $c$ is a constant of order one determined by matching the late-time and early-time solutions.
It is conventional to write the relic density in terms of the Hubble parameter, $h=H_0 /100 {\rm km ~s^{-1}~Mpc^{-1}}$. Observationally, the DM relic abundance is determined to be $\Omega h^2 =0.1186\pm 0.0031$~\cite{Ade:2015xua}.

We plot in Fig. {\ref{density}} the dark matter relic density as the function of dark matter mass.  The red,  blue and green lines correspond, respectively, to $\lambda_\mathrm{eff} =0,$ 2, and 5. 
The  top (bottom) panel gives the septuplet (quintuplet) case.
To obtain the correct relic density, one has $M=9.17~\text{TeV}$ for the septuplet and $M=4.60~\text{TeV}$ for the quintuplet by taking $\lambda_{\rm eff}=0$.

\begin{figure}[t]
\begin{center}
\includegraphics[width=7.5cm,height=7cm,angle=0]{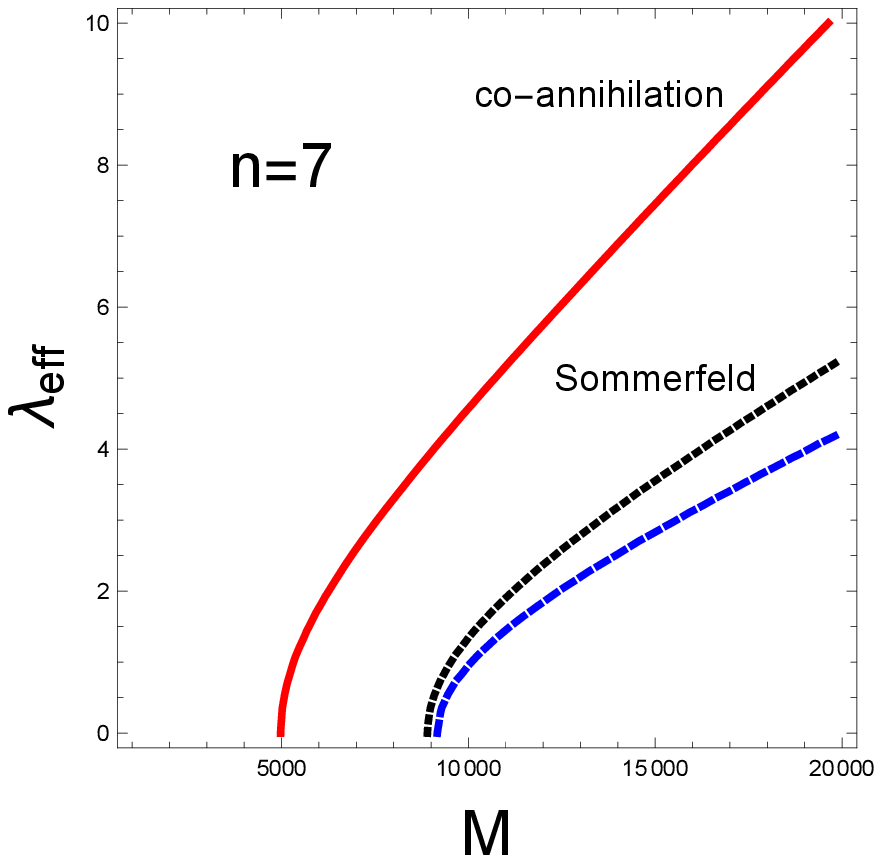}
\includegraphics[width=7.5cm,height=7cm,angle=0]{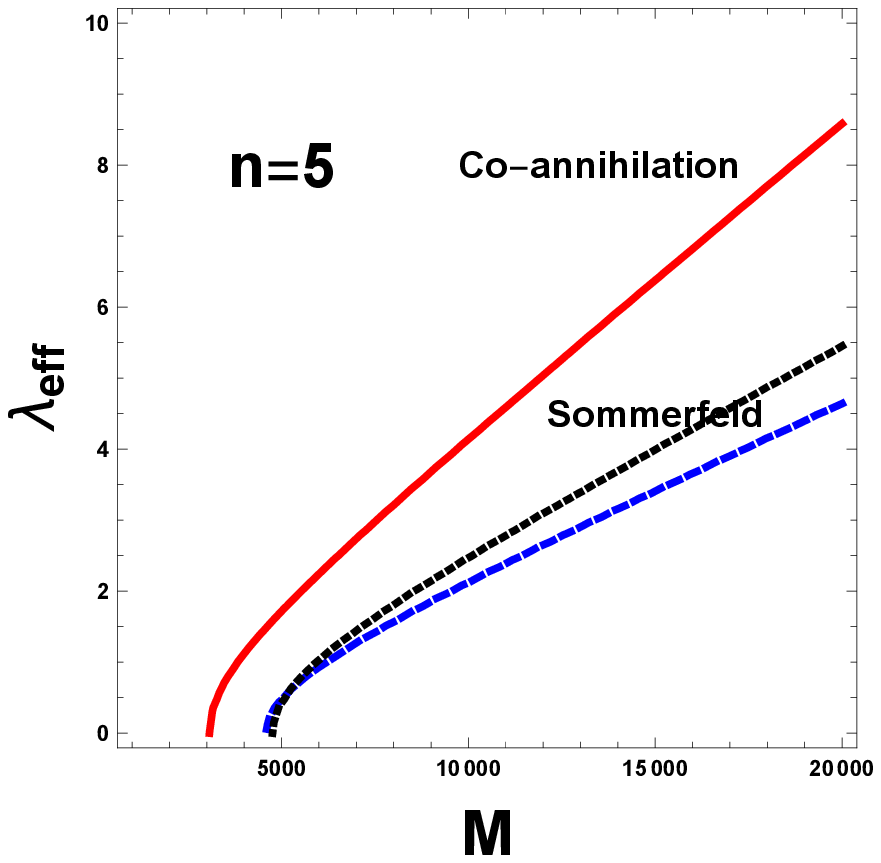}
\end{center}
\caption{ Contours of the dark matter relic density ($\Omega h^2 =0.1186$) in the $M-\lambda_{\rm eff}$ plane. The solid line (red) corresponds to the co-annihilation case, the long-dashed line (blue) represents the one-species scenario.  The short-dashed (black) line includes the Sommerfeld enhancement effect (see Section \ref{sub:sommer} below). The top (bottom) panel describes the septuplet (quintuplet) dark matter case.}  \label{co-anni}\end{figure}

\subsection{Co-annihilation}

The mass splittings between the neutral and charged components of the septuplet is about $166~\text{MeV}$~\cite{Cirelli:2005uq}, so the effect of co-annihilation  should be considered. The relevant processes are listed in Table.~\ref{co-ann}.

\begin{table}[h]
\centering
\begin{tabular}{c|c|c|c|c}
\hline
\hline Process &  & Mediator &  &\\ \hline&  $s-{\rm channel} $ & $t-{\rm channel}$ & u-channel&  ${\rm 4P}  $  \\
\hline
$s^{+Q} s^{-Q} \to W^+W^-$& $h,Z,\gamma$ & $s^{Q-1}$& $s^{-Q+1}$& ${\rm 4P}$ \\
\hline
$s^{+Q} s^{-Q+1}\to W^+ Z(\gamma)$ &$W^+$ &$s^{Q-1}$& $s^{-Q}$ & 4P\\
\hline
$s^{+Q}s^{-Q} \to ZZ (\gamma \gamma, Z\gamma)$ & $h$&$s^Q$ &$s^{-Q}$ & 4P\\
\hline
$s^{+Q} s^{-Q} \to \bar f f $ & $h,W,Z,\gamma$ & & & \\
\hline
$s^{+Q} s^{-Q} \to hh $ & $h$ & $s^{+Q}$ & $s^{-Q} $ & 4P \\
\hline
$s^{+Q} s^{-Q} \to h Z(\gamma)$ & $Z $ & $S^{+Q} $ & $s^{-Q}$ &  \\
\hline
$s^{Q} s^{-Q+1} \to h W^+$ & $W$& $s^{Q}$& $s^{-Q+1}$ & \\
\hline
\hline
\end{tabular}
\caption{ A complete set of process relevant to the co-annihilation of the scalar multiplet dark matter, 4P represents four point interactions.   }\label{co-ann}
\end{table}

The eq.~(\ref{co-annihilation}) can be simplified as~\cite{Edsjo:1997bg}
\begin{eqnarray}
\langle \sigma v_{\rm M\slashed{o}ller}\rangle =  \sum_{ij }   {A_{ij} \over n_{\rm eq}^2 } \label{average}
\end{eqnarray}
where $n_{\rm eq}$ in the denominator is
\begin{eqnarray}
n_{\rm eq} = {T \over 2 \pi^2 } \sum_i g_i m_i^2 K_2 \left( {m_i \over T} \right),
\end{eqnarray}
and $A_{ij}$ in the numerator can be written as
\begin{eqnarray}
A_{ij} = {T \over 64 \pi^4 } \int ds  \sqrt{s}  \beta_{ij } g_i g_j W_{ij}  k_{1} \left(  { \sqrt{s} \over T }\right) \; ,
\end{eqnarray}
with $W_{ij}$ being a dimensionless Lorentz invariant, defined as $W_{ij} = 4 E_i E_j \sigma_{ij} v_{ij}$\footnote{ $v_{ij} $ is defined by $v_{ij} =\sqrt{ (p_i \cdot p_j )^2  -m_i^2 m_j^2 } /E_i E_j$~\cite{Beneke:2014gja}, where $E_i$ and $p_i$ are the Energy of four-momentum of particle $i$. }.

To illustrate the impact of including co-annihilation processes, we plot in Fig.~\ref{co-anni} the value of $\lambda_\mathrm{eff}$ needed to reproduce the observed relic density as a function of the DM mass. The upper (lower) panel corresponds to the septuplet (quintuplet) case. The dashed blue line gives the result for single species annihilation case, while the solid red curve indicates the result including co-annihilation. We observe that the presence of more species initially in equilibrium with the DM requires a larger effective interaction strength to avoid over-saturating the observed relic density. The reason can be seen from Eq.~(\ref{average}), for which the denominator  can be approximated as $n_{\rm eq} \approx (2j+1)^2 n_{{\rm eq}, s}^2$ with $j$ and $n_{{\rm eq}, s}$ being, respectively, the total isospin of the multiplet and the number density of a single component in equilibrium. As $j$ increases, so does $n_{\rm eq}$. On the other hand, the numerator factor, $\sum_{ij} A_{ij}$ only accounts for the combinations of multiplet components that are able to annihilate, and it does not grow as fast as $n_{\rm eq}$ with increasing $j$. Consequently, one must (a) increase $\lambda_\mathrm{eff}$ (for fixed $M$); (b) decrease $M$ (for fixed $\lambda_\mathrm{eff}$); or (c) introduce some combination of both in order to maintain the total cross section as compared to the single species scenario. We refer the reader to Ref.~\cite{Logan:2016ivc} for a similar discussion regarding the $n=6,8$ scalar multiplet dark matter cases.


\subsection{Sommerfeld enhancement }
\label{sub:sommer}

Now we investigate the effect of the non-perturbative electroweak Sommerfeld enhancement~\cite{Hisano:2003ec,ArkaniHamed:2008qn,Slatyer:2009vg}, where the gauge bosons mediate an long-range effective force between the annihilating DM particles.
To that end, we first observe that in the SM, there is no true phase transition between the electroweak symmetric phase and the broken phase, but the cross over is located at $T_c=159\pm1$ GeV~\cite{DOnofrio:2014rug}.
Above this temperature, which can be translated to a critical dark matter mass $M_c\approx 3.2~{\rm TeV}$ (assuming a freeze out temperature set by $x_F\sim 0.05$ with $T_F\sim T_c$) , electroweak symmetry is restored; $W$ and $Z$ bosons can be taken as massless particles; and triple scalar couplings  go to zero as they are proportional to the vaccum expectation value of  neutral component of the Higgs doublet. According to the calculation performed in the last subsection, both the septuplet  and the quintuplet DM  are heavier than $M_c$ for a  sizable $\lambda_{\rm eff}^{} $, so we take the massless gauge boson limit and vanishing triple scalar coupling to evaluate the Sommerfeld enhancement.

Note that we do not consider here the impact of DM-DM bound states, which can lead to an additional enhancement of the annihilation cross section for certain values of $M$. The impact of a bound state on DM annihilation dynamics is most pronounced when the temperature is $\lesssim E_B$, where $E_B$ is the binding energy. As analyzed in detail in Ref.~\cite{Cirelli:2007xd}, however, the Sommerfeld enhancement is plays the most significant role in setting the relic density at temperatures well above $E_B$. Thus, one would expect the presence of the bound states to have a subdominant effect on the overall relic density. Consequently, neglect of the bound state effects appears to be reasonable in the present context.
%
%

To proceed, we consider the Coulomb potential associated with the electroweak gauge bosons is given by~\cite{Strumia:2008cf}
\begin{eqnarray}
V\equiv { a \over r } ={  g^2 \over 32 \pi r  } \left. [(2N+1)^2+1-2n^2 \right.]
\label{eq:sommer1}
\end{eqnarray}
where $N$ is the total isospin of { the  initial state containing two annihilating DM particles}  and $n$ is the dimension of the SU(2$)_L$ irreducible representation of the DM.
Since DM only annihilates into SM final states, one has $N=0,1,2$, depending on the specific process. Of these possibilities, which there exist more $N=0$ final SM final states that those with $N=0,1$, so we concentrate on the $N=0$ case. Note that for $n>1$, the corresponding potential is attractive.

%
{The Sommerfeld enhancement factor $S=\sigma/\sigma_\mathrm{perturbative}$ }
for the Coulomb potential can be written as
\begin{eqnarray}
S= -\pi { a \over \beta} {1 \over 1- \exp({\pi a \over \beta })}
\end{eqnarray}
where $\beta$ is the relative velocity between the annihilating particles  (note that $a<0$ for $N=0$ and $n > 1$).
For a $s$-wave annihilation, one can use the Sommerfeld enhancement averaged over  the thermal distribution, defined as~\cite{Feng:2010zp}
\begin{eqnarray}
\langle  S \rangle  ={ x^{3/2}  \over 2 \sqrt{\pi }} \int S \beta^2 \exp \left( - x  { \beta^2 /4 }\right) d\beta
\end{eqnarray}
where $x=M/T$ with $T$ the temperature.
%

\begin{figure}[t]
\begin{center}
\includegraphics[width=7.5cm,height=7cm,angle=0]{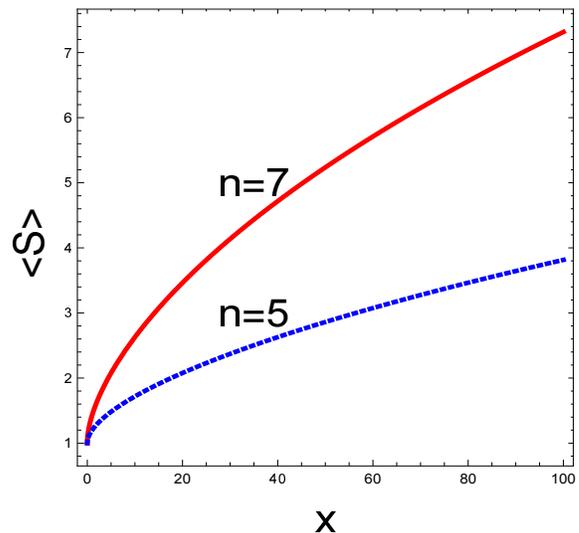}
\end{center}
\caption{Sommerfeld enhancement factor for the quintuplet and septuplet as the function of $x(M/T)$. }  \label{sommerfeld}\end{figure}

%
In Fig.~\ref{sommerfeld} we show the thermal average of the Sommerfeld enhancement as the function of $x$.
A numerical calculation gives $\langle S \rangle \sim 3.4$(septuplet), $2.1$ (quintuplet) at $x=x_F$, which will be used in the calculation of the dark matter relic density.
 As can be seen from Eq.~(\ref{eq:sommer1}), a higher dimensional representation for the multiplet gives rise to a larger enhancement factor.
The resulting impact of the Sommerfeld enhancement  is shown in Fig.~\ref{co-anni}, where the dotted black line corresponds to the case of including both co-annihilation and Sommerfeld enhancement effects. As expected, the presence of this enhancement counteracts the effect of coannihilation, allowing for a smaller value of $\lambda_\mathrm{eff}$ (for fixed $M$) or larger value of $M$ (for fixed $\lambda_\mathrm{eff}$).

\section{Direct detection}
\label{sec:direct}

For conventional Higgs portal dark matter models, constraints from dark matter direct detection are quite severe. The parameter space of these models is strongly constrained by the limits obtained by the LUX~\cite{Akerib:2016vxi}, PandaX-II~\cite{Cui:2017nnn}, and Xenon1T~\cite{Aprile:2018dbl} experiments. In what follows, we consider how the presence of the Higgs portal interactions affects the interpretation of these experimental results.
To that end, we consider all the terms in the effective Lagrangian for low-energy DM interactions with SM particles relevant to the scalar DM scenario considered in this paper.  In the limit $M_{DM} \gg M_W\gg M_q$, one has\cite{Hisano:2011cs,Hisano:2014kua,Hill:2014yka,Hill:2014yxa}
\begin{eqnarray}
{\cal L}_{\rm eff}^{} = {1\over 2 } \lambda_{\rm eff}^{}  {1\over m_h^2}  \Phi_{n,0}^2  \bar q  m_q q + {f_T \over M_\Phi^2 } \Phi_{n,0} (i\partial^\mu) (i \partial^\nu ) \Phi_{n, 0} {\cal O}_{\mu \nu}^q \  \nonumber
\end{eqnarray}
where
\begin{eqnarray}
 {\cal O}_{\mu \nu}^q = {1\over 2 } \bar q i \left( D_\mu \gamma_\nu + D_\nu \gamma_\mu - {1\over 2 } g_{\mu\nu} \slashed{D}\right) q
\end{eqnarray}
is the twist-two quark bilinear with coefficient function\cite{MJRM19}
\begin{widetext}
\begin{eqnarray}
f_T= {\alpha_2^2 \over 8 m_W^2 } {n^2 -(4Y^2 +1) \over 4 } \left\{ \omega \ln \omega +4 + {(4-\omega)(2+\omega) \arctan{2b_\omega/\sqrt{\omega}}\over b_\omega \sqrt{\omega}} \right\}
\label{eq:ft}
\end{eqnarray}
\end{widetext}
and with $\omega=m_W^2 /m_\Phi^2$,  $b_\omega=\sqrt{1-\omega/4}$.

We note that the interaction involving the twist two operator arises from the exchange of two massive electroweak gauge bosons between the DM and quarks inside the nucleus.
We also observe that this contribution differs from what appears in Ref.~\cite{Cirelli:2005uq}, which did not include the effect of the twist-two operator. However, we have confirmed using explicit calculation that the same computation of the two-boson exchange diagrams involving fermionic rather than scalar DM yields the same result as given in Refs.~\cite{Hisano:2011cs,Hisano:2014kua}. To our understanding, the authors of Ref.~\cite{Hambye:2009pw} utilized the expressions in Ref.~\cite{Cirelli:2005uq} when computing the spin-independent direct detection cross section. Consequently, our numerical results given below differ from those of Ref.~\cite{Hambye:2009pw}.

For DM-nucleon scattering, the matrix element can be written as 
\begin{eqnarray}
M_{if} =2m_N^2 \left(f_N^{}    { \lambda_{\rm eff} \over m_h^2 }  + {3\over 4 } f_T^{} f_N^{\rm PDF}\right) \; ,
\end{eqnarray}
where $ f_N^{}  \approx 0.287 (0.284)$~\cite{Belanger:2013oya} for proton(neutron); where  $f_N^{\rm PDF}= 0.526$~\cite{Hisano:2015rsa} is the second moment of the nucleon (proton or neutron) parton distribution function (PDF)  evaluated at $\mu=M_Z^{}$; and where we have taken a normalization appropriate to non-relativistic nuclear states. We note that the expression (\ref{eq:ft}) for  $f_T$ is $\mu$-independent. Inclusion of NLO QCD corrections in the DM-parton scattering amplitude will generate a $\mu$-dependence in $f_T$ that must compensate for the scale dependence of the PDF. We defer a detailed discussion of this feature to Ref.~\cite{MJRM19}. 
The spin-independent cross section then can be written as
\begin{widetext}
\begin{eqnarray}
\sigma_{\rm SI}^{} = { |M_{fi}|^2 \over 16 \pi (m_N + m_\Phi)^2 } = {\mu^2 \over 4 \pi } { m_N^2 \over m_\Phi^2 }\left(f_N^{}    { \lambda_{\rm eff} \over m_h^2 }  + {3\over 4 } f_T^{} f_N^{\rm PDF}\right)^2
\end{eqnarray}
\end{widetext}
where $\mu= m_N^{} M /(m_N^{}+M)$.
\begin{figure}[h]
\begin{center}
\includegraphics[width=7.5cm,height=7cm,angle=0]{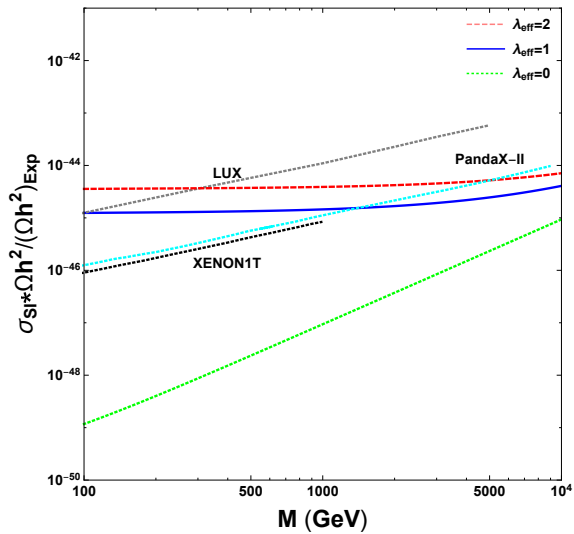}
\end{center}
\caption{DM-nucleus scattering cross section as a function of the dark matter mass for $n=7$. The dashed dashed (red), solid (blue),  and dotted (green) lines correspond to $\lambda_{\rm eff}=$ 2, 1, 0 respectively. The gray, cyan and black dotted lines give the exclusion limits of LUX, PandaX-II and XENON1T respectively. }  \label{direct}\end{figure}

In Fig. \ref{direct}, we plot as a function of $M$ the cross section of the DM-proton cross section, scaled by the fraction of the relic density corresponding to the value of $M$ as obtained in our computation of Section \ref{sec:relic}.
Taking the septuplet for illustration, the dashed dashed (red), solid (blue),  and dotted (green) lines correspond to $\lambda_{\rm eff}=$ 2, 1, 0 respectively.
%
%
{The gray, cyan and black dotted lines give the exclusion limits of LUX, PandaX-II and XENON1T respectively.
We observe that the Higgs portal interactions dominate the scaled spin-independent cross section for a sizable $\lambda_{\rm eff}$. The contribution of twist-2 effective operator, indicated by the $\lambda_{\rm eff}=0$ curve, becomes relatively sizable only for heavy DM, though its impact still lies well below the sensitivity of the present direct detection experiments.
The situation is different in the evaluation of the relic abundance, where the gauge interactions dominate the annihilation.
As a result, one can easily find the parameter space that may give rise to an observed relic abundance and a small direct detection cross section. Conversely, including the effects of both co-annihilation and the Sommerfeld enhancement, we observe that saturating the observed relic density and evading the present direct detection limits require a rather small value of $|\lambda_{\rm eff}|$. To illustrate, consider the septuplet case. From Fig.~\ref{density} we see that obtaining the relic density requires $M$ in the vicinity of 9 TeV for vanishing $\lambda_{\rm eff}$. On the other hand, for $\lambda_{\rm eff}=1$, the present direct detection results constrain $M$ to be no larger than about one TeV -- a value for which the fraction of the relic density would lie well below the observed value. Looking ahead to next generation direct detection experiments and assuming that the only thermal WIMP is the neutral component of the septuplet, we conclude that the observation of a non-zero signal would likely require the presence of a significant, non-zero $\lambda_{\rm eff}$. In this case, the septuplet would comprise at most only a modest fraction of the relic density, with the remaining corresponding to a non-thermal and/or non-WIMP species.

\section{Conclusions}

In this paper we have revisited earlier analyses of scalar electroweak multiplet dark matter. After presenting the most general, renormalizable potential for a electroweak multiplet $\Phi$ that interacts with the SM Higgs doublet, we show that in general the Higgs portal coupling depends on three independent parameters in the potential. In order to ensure that the neutral component of $\Phi$ yields the lowest mass state, ensuring its viability as a DM candidate, one of these couplings must be vanishingly small. The resulting dynamics of DM annihilation and DM-nucleus scattering then depend on a single effective coupling, $\lambda_\mathrm{eff}$. After evaluating the DM relic abundance by considering effects of both co-annihilation and Sommerfeld enhancement, we calculated for the first time the spin-independent direct detection cross section by taking into account the contribution of the twist-2 effective operators, which turns to be important for a heavy scalar DM. Focusing on the electroweak quintuplet and septuplet for illustration, we find that for $\lambda_\mathrm{eff}\sim\mathcal{O}(1)$ present DM direct detection limits imply that the electroweak multiplet mass scale $M$ most be $\lesssim 1$ TeV. In this case, the neutral electroweak multiplet scalar would comprise a subdominant component of the DM relic density. }

\label{sec:conclude}

\begin{acknowledgments}
{ WC was supported in part by the Natural Science Foundation of China
under Grant No. 11775025 and the Fundamental Research Funds for the Central Universities. {GJD was supported in part by the support of the National Natural Science Foundation of China under Grant No 11522546}.
XGH was supported in part by the MOST (Grant No. MOST 106-2112-M-002-003-MY3 ), and in part by Key Laboratory for Particle Physics,
Astrophysics and Cosmology, Ministry of Education, and Shanghai Key Laboratory for Particle
Physics and Cosmology (Grant No. 15DZ2272100), and in part by the NSFC (Grant Nos. 11575111 and 11735010).
MJRM was supported in part under U.S. Department of Energy contract DE-SC0011095.}
\end{acknowledgments}

\vskip 0.2in

\appendix

\section{$SU(2)$ group theory}
\label{sec:groupA}

The Lie algebra of the $SU(2)$ group is specified by
\begin{equation}
\left[J_i,J_{j}\right]=i\epsilon_{ijk}J_{k},\qquad i,j,k=1,2,3\,.
\end{equation}
$SU(2)$ has only one Casimir operator
\begin{equation}
J^2=J^2_1+J^2_2+J^2_3\,.
\end{equation}
We can define familiar raising and lowering operators:
\begin{equation}
J_{\pm}=J_1\pm iJ_2\,.
\end{equation}
They satisfy the following commutation relation
\begin{equation}
\left[J_3,J_{\pm}\right]=\pm J_{\pm}\,.
\end{equation}
The eigenstate $|j,m\rangle$ can be labelled by the eigenvalues of $J^2$ and $J_3$:
\begin{eqnarray}
\nonumber&&J^2|j,m\rangle=j(j+1)|j,m\rangle,\\
&&J_3|j,m\rangle=m|j,m\rangle\,,
\end{eqnarray}
where $j$ can be any half integer, and $m=-j, -j+1,\ldots, j-1, j$. The different states within a multiplet can be generated by acting with the  raising and lowering operators,
\begin{equation}
J_{\pm}|j,m\rangle=\sqrt{(j\mp m)(j\pm m+1)}|j,m\pm1\rangle
\end{equation}
Consequently we have
\begin{eqnarray}
\nonumber&&\langle j,m^{\prime}|J_{+}|j,m \rangle=\sqrt{(j-m)(j+m+1)}\;\delta_{m^{\prime},m+1},\\
&&\langle j,m^{\prime}|J_{-}|j,m \rangle=\sqrt{(j+m)(j-m+1)}\;\delta_{m^{\prime},m-1}
\end{eqnarray}
We can form a $2j+1$ representation by choosing the following $2j+1$ orthogonal states as base vectors:
{\small
\begin{equation}
\left(\begin{array}{c}
1\\
0\\
0\\
.\\
.\\
.\\
0\\
0
\end{array}\right)\equiv|j,j\rangle,~ \left(\begin{array}{c}
0\\
1\\
0\\
.\\
.\\
.\\
0\\
0
\end{array}\right)\equiv|j,j-1\rangle,\cdots~ \left(\begin{array}{c}
0\\
0\\
0\\
.\\
.\\
.\\
0\\
1
\end{array}\right)\equiv|j,-j\rangle\,.
\end{equation}
}
The representation matrices for the generators $J_{+}$, $J_{-}$ and $J_{3}$ are

\begin{widetext}
\begin{eqnarray}
&&J_{+}=\left(\begin{array}{ccccccc}
0  &  \sqrt{2j}  & 0  & 0 &    \ldots  &  0  &  0 \\
0  &  0  &  \sqrt{2(2j-1)}  &  0  &  \ldots  &  0  &  0  \\
0  & 0  &  0  & \sqrt{6(j-1)}   &  \ldots  &  0  &  0  \\
\ldots  &  \ldots  &  \ldots  &  \ldots   &  \ldots  &  \ldots  &  \ldots   \\
0  &  0 &  0  &  0  & \ldots & \sqrt{2(2j-1)}  & 0  \\
0  &  0 &  0  &  0 & \ldots  &  0  & \sqrt{2j} \\
0  &  0 &  0  &  0  & \ldots  &  0  & 0
\end{array}\right),\\
&&J_{-}=\left(\begin{array}{ccccccc}
0 &  0  &  0 & \ldots & 0 &  0  &  0 \\
\sqrt{2j}  & 0  &  0 &  \ldots & 0 &  0  &  0 \\
0  & \sqrt{2(2j-1)}  &   0  & \ldots & 0 &  0  &  0 \\
0  & 0  &  \sqrt{6(j-1)} &  \ldots & 0 &  0  &  0 \\
\ldots  & \ldots & \ldots & \ldots & \ldots & \ldots  &  \ldots \\
0  & 0  & 0 & \ldots &  \sqrt{2(2j-1)} & 0  & 0 \\
0  & 0  & 0 & \ldots &  0 & \sqrt{2j}  & 0
\end{array}
\right),\\
&&J_{3}=\left(\begin{array}{cccccc}
j  ~&~ 0 ~&~  0 ~&~ \ldots  ~&~ 0  ~&~ 0\\
0 ~&~  j-1 ~&~ 0 ~&~ \ldots ~&~ 0  ~&~ 0\\
0  ~&~  0 ~&~ j-2 ~&~ \ldots  ~&~ 0  ~&~ 0\\
\ldots ~&~  \ldots ~&~ \ldots ~&~ \ldots ~&~ \ldots  ~&~ \ldots \\
0  ~&~  0 ~&~ 0 ~&~ \ldots  ~&~ -j+1 ~&~ 0\\
0  ~&~  0 ~&~ 0 ~&~ \ldots  ~&~ 0 ~&~ -j
\end{array}
\right)
\end{eqnarray}
\end{widetext}
The representation matrix for each group element of $SU(2)$ can be expressed as
\begin{equation}
\exp\left(i\sum_{k=1}^{3}\alpha_kJ_{k}\right)\,,
\end{equation}
where $\alpha_{k}(k=1,2,3)$ are real parameters and
\begin{equation}
J_{1}=\frac{1}{2}\left(J_{+}+J_{-}\right),\qquad J_2=-\frac{i}{2}\left(J_{+}-J_{-}\right)\,.
\end{equation}
It is well-known that $SU(2)$ has a unique irreducible representation for each spin $j$. Hence each representation should be equivalent to its complex conjugate representation. We find the unitary transformation relating representation and its complex conjugate is
\begin{equation}
V=\left(\begin{array}{ccccc}
0  ~&~  0  ~&~  \ldots ~&~ 0 ~&~ 1 \\
0 ~&~  0  ~&~  \ldots ~&~ -1 ~&~ 0 \\
\ldots  ~&~  \ldots  ~&~  \ldots ~&~ \ldots ~&~ \ldots \\
0  ~&~  (-1)^{2j-1}  ~&~  \ldots ~&~ 0 ~&~  0\\
(-1)^{2j}  ~&~ 0   ~&~  \ldots ~&~ 0 ~&~  0
\end{array}
\right)\,,
\end{equation}
which fulfills $V_{ik}=(-1)^{i+1}\delta_{i+k,2j+2}$. Note that the unitary transformation $V$ reduces to the familiar form for $j=\frac{1}{2}$,
\begin{equation}
V=\left(\begin{array}{cc}
0  & 1 \\
-1  & 0
\end{array}
\right),\qquad \text{for}\quad j=\frac{1}{2}\,.
\end{equation}
One can easily check that
\begin{equation}
-VJ^{*}_{\pm}V^{-1}=J_{\mp},\qquad -VJ^{*}_{3}V^{-1}=J_{3}\,,
\end{equation}
which leads to
\begin{equation}
-VJ^{*}_{k}V^{-1}=J_{k},\quad k=1,2,3\,.
\end{equation}
Consequently we have
\begin{equation}
V\left(\exp\left(i\sum_{k=1}^{3}\alpha_kJ_{k}\right)\right)^{*}V^{-1}=\exp\left(i\sum_{k=1}^{3}\alpha_kJ_{k}\right)\,,
\end{equation}
which implies each representation and its complex conjugate are really equivalent, and the similarity transformation is indeed given by $V$. As a result, for a $SU(2)$ multiplet $\Phi$ in the representation $j$ with
\begin{equation}
\Phi=\left(\begin{array}{c}
\phi_{j,j}\\
\phi_{j,j-1}\\
\phi_{j,j-2}\\
.\\
.\\
.\\
\phi_{j,-j+1}\\
\phi_{j,-j}\\
\end{array}
\right)\,,
\end{equation}
where the subscript denotes the eigenvalues of $J^2$ and $J_3$. The state $\overline{\Phi}$ would transform in the same way as $\Phi$ with \begin{equation}
\overline{\Phi}=V\Phi^{*}=\left(\begin{array}{c}
\phi^{*}_{j,-j}\\
-\phi^{*}_{j,-j+1}\\
\phi^{*}_{j,-j+2}\\
.\\
.\\
.\\
(-1)^{2j-1}\phi^{*}_{j,j-1}\\
(-1)^{2j}\phi^{*}_{j,j}\\
\end{array}
\right)\,.
\end{equation}
Note that it is very convenient to construct $SU(2)$ invariant from $\overline{\Phi}$ instead of $\Phi^{*}$.

\section{The renormalizable scalar potential of Higgs and a scalar multiplet}
\label{sec:groupB}

If we extend the standard model by introducing a scalar electroweak multiplet $\Phi$ of isospin $j$, the one-loop beta function of $SU(2)$ gauge coupling for the Standard Model would be modified into
\begin{equation}
\beta(g)=\frac{g^3}{16\pi^2}\big[-\frac{19}{6}+\frac{1}{9}j(j+1)(2j+1)\big]\,.
\end{equation}
We can see that $\beta(g)$ remains negative only for $j\leq\frac{3}{2}$. For $j\geq 2$, it becomes positive and hits the Landau pole. For instance adding a scalar multiplet with isospin $j\geq 5$ will bring the Landau pole of $SU(2)$ gauge coupling at $\Lambda\leq 10$ TeV and it is even smaller $\Lambda\leq 180$ GeV for $j\geq 10$. Therefore, perturbativity of gauge coupling at the TeV scale constraints the isospin of the multiplet to be $j\leq 5$.

Another bound on the size of a electroweak multiplet is set by perturbative unitarity of tree-level scattering
amplitude. In Ref.~\cite{Hally:2012pu,Earl:2013jsa}, the $2\rightarrow 2$ scattering amplitudes for
scalar pair annihilations into electroweak gauge bosons have been
computed and by requiring zeroth partial wave amplitude satisfying the
unitarity bound, it was shown that maximum allowed complex $SU(2)$
multiplet would have isospin $j\leq 7/2$ and real multiplet would have
$j\leq 4$. In the following, we shall report the most general renormalizable scalar potential $V(\Phi)$ for $\Phi$ and the interaction potential $V(\Phi, H)$ between $\Phi$ and $H$. The  hypercharge of $\Phi$ is denoted by $Y$.

\subsection{Integer isospin $j$ }

The electroweak multiplet $\Phi$ has $2j+1$ component fields, and the coupling of each component of $\Phi$ to the $Z$ boson is proportional to $T_3-Q\sin^2\theta_{W}$ with the electric charge $Q=T_3+Y/2$. If the hypercharge is nonzero $Y\neq0$, the neutral component of $\Phi$ has unsuppressed vector interaction with $Z$ such that it can not be dark matter candidate because of the constraints from direct detection. On the other hand, for $Y=0$, the neutral component of $\Phi$ could be potential dark matter candidate. The scalar multiplet $\Phi$ can be real or complex. If $\Phi$ is a real multiplet, there is a redundancy $\Phi=\overline{\Phi}$ such that the constraint $\phi_{j,m}=(-1)^{j-m}\phi^{*}_{j,-m}$ should be fulfilled. For complex multiplet, each component represents a unique field, and it can be decomposed into two real multiplets as follows
\begin{equation}
A=\frac{1}{\sqrt{2}}\left(\Phi+\overline{\Phi}\right),\qquad B=\frac{i}{\sqrt{2}}\left(\Phi-\overline{\Phi}\right)\,.
\end{equation}
It is easy to check that both $A$ and $B$ fulfill the realness condition $A=\overline{A}$ and $B=\overline{B}$. Therefore a
general model with a complex multiplet $\Phi$ is equivalent to a model of two interacting real multiplets $A$ and $B$. We shall present the concrete form of the scalar potentials $V(\Phi)$ and $V(H, \Phi)$ for different cases of $Y=0$ and $Y\neq0$.
\begin{itemize}[labelindent=-0.4em, leftmargin=1.6em]
\item{Complex $\Phi$ with $Y\neq0$ and $Y\neq\pm2$}
\begin{eqnarray}
\nonumber V(\Phi)&=&M^2_{\Phi}\Phi^{\dagger}\Phi+\sum^{2j}_{\mathcal{J}=0}\lambda_{\mathcal{J}}[(\Phi\Phi)_{\mathcal{J}}(\overline{\Phi}\;\overline{\Phi})_{\mathcal{J}}]_{0}\,,\\
V(H, \Phi)&=&\alpha(H^{\dagger}H)(\Phi^{\dagger}\Phi)+\beta[(\overline{H}H)_{1}(\overline{\Phi}\Phi)_{1}]_{0},
\end{eqnarray}
where only the terms of even $\mathcal{J}$ lead to nonzero contribution, $\overline{H}=i\sigma_{2}H^{\ast}$ with $\sigma_2$ being Pauli matrix. The contraction $[(\Phi\Phi)_{\mathcal{J}}(\overline{\Phi}\,\overline{\Phi})_{\mathcal{J}}]_{0}$ is given by
{\small\begin{equation}
\label{eq:quartic_integer}[(\Phi\Phi)_{\mathcal{J}}(\overline{\Phi}\,\overline{\Phi})_{\mathcal{J}}]_{0}
=\sum_{m}C^{0,0}_{\mathcal{J},m;\mathcal{J},-m}(\Phi\Phi)_{\mathcal{J},m}(\overline{\Phi}\,\overline{\Phi})_{\mathcal{J},-m}\,,
\end{equation}}
with
\begin{eqnarray}
\nonumber(\Phi\Phi)_{\mathcal{J},m}&=&\sum_{m_1,m_2}C^{\mathcal{J},m}_{j,m_1;j,m_2}\phi_{j,m_1}\phi_{j,m_2},\\
\nonumber(\overline{\Phi}\;\overline{\Phi})_{\mathcal{J},m}&=&\sum_{m_1,m_2}(-1)^{2j-m_1-m_2}\\
\label{eq:HH_contr_general}&&\times C^{\mathcal{J},m}_{j,m_1;j,m_2}\phi^{*}_{j,-m_1}\phi^{*}_{j,-m_2}\,.
\end{eqnarray}
Consequently the contraction $(\Phi\Phi)_{\mathcal{J}}$ vanishes for odd $\mathcal{J}$. Notice that all the independent self interactions of $\Phi$ are included here while only two terms are considered in~\cite{Hambye:2009pw,AbdusSalam:2013eya}.

\item{Complex $\Phi$ with $Y=2$ }
\begin{eqnarray}
\nonumber V(\Phi)&=&M^2_{\Phi}\Phi^{\dagger}\Phi+\sum^{2j}_{\mathcal{J}=0}\lambda_{\mathcal{J}}[(\Phi\Phi)_{\mathcal{J}}(\overline{\Phi}\,\overline{\Phi})_{\mathcal{J}}]_{0}\,,\\
\nonumber V(H, \Phi)&=&\alpha(H^{\dagger}H)(\Phi^{\dagger}\Phi)+\beta[(\overline{H}H)_{1}(\overline{\Phi}\Phi)_{1}]_{0}\\
&&+[\mu((HH)_{1}\overline{\Phi})_{0}\delta_{j,1}+\textrm{h.c.}]\,,
\end{eqnarray}
We see that an additional term $[(HH)_{1}\overline{\Phi}]_{0}$ and its hermitian conjugate are allowed if $\Phi$ is a isospin triplet with $j=1$ and $Y=2$. This term would disappear if one adopts a $Z_2$ symmetry under which all SM particles are $Z_2$ even and extra scalar $\Phi$ is $Z_2$ odd.

\item{Complex $\Phi$ with $Y=-2$}
\begin{eqnarray}
\nonumber V(\Phi)&=&M^2_{\Phi}\Phi^{\dagger}\Phi+\sum^{2j}_{\mathcal{J}=0}\lambda_{\mathcal{J}}[(\Phi\Phi)_{\mathcal{J}}(\overline{\Phi}\,\overline{\Phi})_{\mathcal{J}}]_{0}\,,\\
\nonumber V(H, \Phi)&=&\alpha(H^{\dagger}H)(\Phi^{\dagger}\Phi)+\beta[(\overline{H}H)_{1}(\overline{\Phi}\Phi)_{1}]_{0}\\
&&+\left[\mu\left(\left(HH\right)_{1}\Phi\right)_{0}\delta_{j,1}+\textrm{h.c.}\right]\,,
\end{eqnarray}

\item{Complex $\Phi$ with $Y=0$}
{\small\begin{eqnarray}
\nonumber&& V(\Phi)=M^2_{\Phi}\Phi^{\dagger}\Phi+\big[M^{\prime2}_{\Phi}(\Phi\Phi)_{0}\textrm{+h.c.}\big]\\
\nonumber&&\quad+\delta_{0,j\hskip-0.1in\pmod{2}}\big[\mu_1(\Phi(\Phi\Phi)_{j})_{0}+\mu_2(\overline{\Phi}(\Phi\Phi)_{j})_{0}+\textrm{h.c.}\big]\\
\nonumber&&\quad+\sum^{2j}_{\mathcal{J}=0}\lambda_{\mathcal{J}}[(\Phi\Phi)_{\mathcal{J}}(\overline{\Phi}\,\overline{\Phi})_{\mathcal{J}}]_{0} +\sum^{2j}_{\mathcal{K}=0}\Big\{\lambda^{\prime}_{\mathcal{K}}[(\Phi\Phi)_{\mathcal{K}}(\Phi\Phi)_{\mathcal{K}}]_{0}\\
&&\quad
+\lambda^{\prime\prime}_{\mathcal{K}}[(\overline{\Phi}\Phi)_{\mathcal{K}}(\Phi\Phi)_{\mathcal{K}}]_{0}+\textrm{h.c.}\Big\}\,,\\
\nonumber&& V(H, \Phi)=\big[\mu_3(\overline{H}H)_{0}\Phi\delta_{j,0}+\mu_4((\overline{H}H)_{1}\Phi)_{0}\delta_{j,1}+\textrm{h.c.}\big]\\
\nonumber&&\quad+\alpha(H^{\dagger}H)(\Phi^{\dagger}\Phi)+\beta[(\overline{H}H)_{1}(\overline{\Phi}\Phi)_{1}]_{0}\\
&&\quad+\left[\gamma(\overline{H}H)_{0}(\Phi\Phi)_{0}+\textrm{h.c.}\right]\,.
\end{eqnarray}}
Notice that not all the interaction terms $[(\Phi\Phi)_{\mathcal{K}}](\Phi\Phi)_{\mathcal{K}}]_{0}$ for $\mathcal{K}=0, 2, \ldots, 2j$ are independent from each other. For $j=0, 1, 2$, there is only one independent interaction $(\Phi\Phi)_{0}(\Phi\Phi)_{0}$. We find two independent contractions $(\Phi\Phi)_{0}(\Phi\Phi)_{0}$ and $[(\Phi\Phi)_{2}(\Phi\Phi)_{2}]_{0}$ for $j=3, 4, 5$. However, there are four independent interaction terms $(\Phi\Phi)_{0}(\Phi\Phi)_{0}$, $[(\Phi\Phi)_{2}(\Phi\Phi)_{2}]_{0}$,
$[(\Phi\Phi)_{4}(\Phi\Phi)_{4}]_{0}$ and $[(\Phi\Phi)_{6}(\Phi\Phi)_{6}]_{0}$ in the case of $j=10$.  For any given isospin $j$, we can straightforwardly find all the independent contractions among $[(\Phi\Phi)_{\mathcal{K}}(\Phi\Phi)_{\mathcal{K}}]_{0}$ with $\mathcal{K}=0, 2, \ldots, 2j$. The same holds true for the contractions $[(\overline{\Phi}\Phi)_{\mathcal{K}}(\Phi\Phi)_{\mathcal{K}}]_{0}$.

\item{Real $\Phi$ with $Y=0$}
\begin{eqnarray}
\nonumber V(\Phi)&=&\frac{1}{2}M^2_{\Phi}\Phi^{\dagger}\Phi+\mu_1[\Phi(\Phi\Phi)_{j}]_{0}\delta_{0,j\hskip-0.1in\pmod{2}}\\
\nonumber&&+\sum^{2j}_{\mathcal{K}=0}\lambda_{\mathcal{K}}[(\Phi\Phi)_{\mathcal{K}}(\Phi\Phi)_{\mathcal{K}}]_{0}\\
\nonumber V(H, \Phi)&=&\mu_2(\overline{H}H)_{0}\Phi\delta_{j,0}+\mu_3[(\overline{H}H)_{1}\Phi]_{0}\delta_{j,1}\\
&&+\alpha(\overline{H}H)_{0}(\Phi\Phi)_{0}\,.
\end{eqnarray}
As regards the quartic self interaction terms $[(\Phi\Phi)_{\mathcal{K}}(\Phi\Phi)_{\mathcal{K}}]_{0}$ with $\mathcal{K}=0, 2, \ldots, 2j$, there is only one independent contraction $(\Phi\Phi)_{0}(\Phi\Phi)_{0}$ for $j=0, 1, 2$. We find two independent contractions $(\Phi\Phi)_{0}(\Phi\Phi)_{0}$ and $[(\Phi\Phi)_{2}(\Phi\Phi)_{2}]_{0}$ for the case of $j=3, 4, 5$.

\end{itemize}

\subsection{Half integer isospin $j$}

A scalar multiplet $\Phi$ of half integer isospin is always complex for any value of hypercharge $Y$. In other words, the realness condition $\Phi=\overline{\Phi}$ can not be fulfilled anymore.

\begin{itemize}[labelindent=-0.4em, leftmargin=1.6em]

\item{Generic $\Phi$ with $Y\neq0$, $Y\neq\pm1$ and $Y\neq\pm3$}
\begin{eqnarray}
\nonumber V(\Phi)&=&M^2_{\Phi}\Phi^{\dagger}\Phi+\sum^{2j}_{\mathcal{J}=1}\lambda_{\mathcal{J}}[(\Phi\Phi)_{\mathcal{J}}(\overline{\Phi}\,\overline{\Phi})_{\mathcal{J}}]_{0}\,,\\
V(H, \Phi)&=&\alpha(H^{\dagger}H)(\Phi^{\dagger}\Phi)+\beta[(\overline{H}H)_{1}(\overline{\Phi}\Phi)_{1}]_{0}\,,
\end{eqnarray}
where $\mathcal{J}$ should be odd otherwise the contraction $(\Phi\Phi)_{\mathcal{J}}$ is vanishing.

\item{$\Phi$ with $Y=0$}
\begin{eqnarray}
\nonumber&& V(\Phi)=M^2_{\Phi}\Phi^{\dagger}\Phi+\sum^{2j}_{\mathcal{J}=1}\lambda_{\mathcal{J}}[(\Phi\Phi)_{\mathcal{J}}(\overline{\Phi}\,\overline{\Phi})_{\mathcal{J}}]_{0}\\
\nonumber&&~~+\sum^{2j}_{\mathcal{K}=1}\Big\{\lambda^{\prime}_{\mathcal{K}}[(\Phi\Phi)_{\mathcal{K}}(\Phi\Phi)_{\mathcal{K}}]_{0}
+\lambda^{\prime\prime}_{\mathcal{K}}[(\overline{\Phi}\Phi)_{\mathcal{K}}(\Phi\Phi)_{\mathcal{K}}]_{0}+\mathrm{h.c.}\Big\}\,,\\
\nonumber&&V(H, \Phi)=\alpha(H^{\dagger}H)(\Phi^{\dagger}\Phi)+\beta[(\overline{H}H)_{1}(\overline{\Phi}\Phi)_{1}]_{0}\\
&&~~+\big\{\gamma[(\overline{H}H)_{1}(\Phi\Phi)_{1}]_{0}+\mathrm{h.c.}\big\}\,,
\end{eqnarray}
Both interaction terms $[(\Phi\Phi)_{\mathcal{K}}(\Phi\Phi)_{\mathcal{K}}]_{0}$ and $[(\overline{\Phi}\Phi)_{\mathcal{K}}(\Phi\Phi)_{\mathcal{K}}]_{0}$ are vanishing for $j=1/2$. There is only one independent contraction $[(\Phi\Phi)_{1}(\Phi\Phi)_{1}]_{0}$ for $j=3/2, 5/2, 7/2$. We find only two independent terms $[(\Phi\Phi)_{1}(\Phi\Phi)_{1}]_{0}$ and $[(\Phi\Phi)_{3}(\Phi\Phi)_{3}]_{0}$ in case of $j=9/2$.

\item{$\Phi$ with $Y=1$}
\begin{eqnarray}
\nonumber V(\Phi)&=&M^2_{\Phi}\Phi^{\dagger}\Phi+\sum^{2j}_{\mathcal{J}=1}\lambda_{\mathcal{J}}[(\Phi\Phi)_{\mathcal{J}}(\overline{\Phi}\,\overline{\Phi})_{\mathcal{J}}]_{0}\,,\\
\nonumber V(H, \Phi)&=&\alpha(H^{\dagger}H)(\Phi^{\dagger}\Phi)+\beta[(\overline{H}H)_{1}(\overline{\Phi}\Phi)_{1}]_{0}\\
\nonumber&&\hskip-0.3in+\Big\{\gamma_1[(H\Phi)_{j+\frac{1}{2}}(\overline{\Phi}\,\overline{\Phi})_{j+\frac{1}{2}}]_{0}\,\delta_{0,j-\frac{1}{2}\hskip-0.1in\pmod{2}}\\
\nonumber&&\hskip-0.3in+\gamma_2[(H\Phi)_{j-\frac{1}{2}}(\overline{\Phi}\,\overline{\Phi})_{j-\frac{1}{2}}]_{0}\,\delta_{0,j+\frac{1}{2}\hskip-0.1in\pmod{2}}+\textrm{h.c.}\Big\}\\
\nonumber&&\hskip-0.3in+\Big\{\kappa_1[(HH)_{1}(\overline{\Phi}\,\overline{\Phi})_{1}]_{0}+\kappa_2[(HH)_{1}(\overline{H}\,\overline{\Phi})_{1}]_{0}\,\delta_{j,\frac{1}{2}}\\
&&\hskip-0.3in+\kappa_3[(HH)_{1}(\overline{H}\,\overline{\Phi})_{1}]_{0}\,\delta_{j,\frac{3}{2}}+h.c.\Big\}.
\end{eqnarray}

\item{$\Phi$ with $Y=-1$}
\begin{eqnarray}
\nonumber V(\Phi)&=&M^2_{\Phi}\Phi^{\dagger}\Phi+\sum^{2j}_{\mathcal{J}=1}\lambda_{\mathcal{J}}[(\Phi\Phi)_{\mathcal{J}}(\overline{\Phi}\,\overline{\Phi})_{\mathcal{J}}]_{0}\,,\\
\nonumber V(H, \Phi)&=&\alpha(H^{\dagger}H)(\Phi^{\dagger}\Phi)+\beta[(\overline{H}H)_{1}(\overline{\Phi}\Phi)_{1}]_{0}\\
\nonumber&&\hskip-0.3in+\Big\{\gamma_1[(H\overline{\Phi})_{j+\frac{1}{2}}(\Phi\,\Phi)_{j+\frac{1}{2}}]_{0}\,\delta_{0,j-\frac{1}{2}\hskip-0.1in\pmod{2}}\\
\nonumber&&\hskip-0.3in+\gamma_2[(H\overline{\Phi})_{j-\frac{1}{2}}(\Phi\,\Phi)_{j-\frac{1}{2}}]_{0}\,\delta_{0,j+\frac{1}{2}\hskip-0.1in\pmod{2}}+\mathrm{h.c.}\Big\}\\
\nonumber&&\hskip-0.3in+\Big\{\kappa_1[(HH)_{1}(\Phi\,\Phi)_{1}]_{0}+\kappa_2[(HH)_{1}(\overline{H}\,\Phi)_{1}]_{0}\,\delta_{j,\frac{1}{2}}\\
&&\hskip-0.3in+\kappa_3[(HH)_{1}(\overline{H}\,\Phi)_{1}]_{0}\,\delta_{j,\frac{3}{2}}+\mathrm{h.c.}\Big\}
\end{eqnarray}

\item{$\Phi$ with $Y=3$}
\begin{eqnarray}
\nonumber V(\Phi)&=&M^2_{\Phi}\Phi^{\dagger}\Phi+\sum^{2j}_{\mathcal{J}=1}\lambda_{\mathcal{J}}[(\Phi\Phi)_{\mathcal{J}}(\overline{\Phi}\,\overline{\Phi})_{\mathcal{J}}]_{0}\,,\\
\nonumber V(H, \Phi)&=&\alpha(H^{\dagger}H)(\Phi^{\dagger}\Phi)+\beta[(\overline{H}H)_{1}(\overline{\Phi}\Phi)_{1}]_{0}\\
&&\hskip-0.3in+\delta_{j, \frac{3}{2}}\big\{\gamma[(HH)_{1}(H\overline{\Phi})_{1}]_{0}+\mathrm{h.c.}\big\}.
\end{eqnarray}

\item{$\Phi$ with $Y=-3$}
\begin{eqnarray}
\nonumber V(\Phi)&=&M^2_{\Phi}\Phi^{\dagger}\Phi+\sum^{2j}_{\mathcal{J}=1}\lambda_{\mathcal{J}}[(\Phi\Phi)_{\mathcal{J}}(\overline{\Phi}\,\overline{\Phi})_{\mathcal{J}}]_{0}\,,\\
\nonumber V(H, \Phi)&=&\alpha(H^{\dagger}H)(\Phi^{\dagger}\Phi)+\beta[(\overline{H}H)_{1}(\overline{\Phi}\Phi)_{1}]_{0}\\
&&\hskip-0.3in+\delta_{j, \frac{3}{2}}\big\{\gamma[(HH)_{1}(H\Phi)_{1}]_{0}+\mathrm{h.c.}\big\}.
\end{eqnarray}

\end{itemize}

\section{Self-Interactions}
\label{sec:groupC}

Starting from Eq.~(\ref{eq:selfquartic}) a direct calculation yields the self interactions among the neutral fields $\phi_{3; (0, +)}$ and $\phi_{3; (0,-)}$

\begin{eqnarray}
\nonumber&&\frac{1}{4}\left[\tilde{\kappa}_1+2\mathrm{Re}(\tilde{\kappa}_2)+2\mathrm{Re}(\tilde{\kappa}_3)\right]\phi_{3;(0,+)}^4\\
&&-\left[2\mathrm{Im}(\tilde{\kappa}_2)+\mathrm{Im}(\tilde{\kappa}_3)\right]\phi_{3;(0,+)}^3\phi_{3;(0,-)}\\
\nonumber&&+\frac{1}{2}\left[\tilde{\kappa}_1-6\mathrm{Re}(\tilde{\kappa}_2)\right]\phi_{3;(0,+)}^2\phi_{3;(0,-)}^2\\
\nonumber&&+\left[2\mathrm{Im}(\tilde{\kappa}_2)-\mathrm{Im}(\tilde{\kappa}_3)\right]\phi_{3;(0,+)}\phi_{3;(0,-)}^3\\
\label{eq:interactions_ab}&&+\frac{1}{4}\left[\tilde{\kappa}_1+2\mathrm{Re}(\tilde{\kappa}_2)-2\mathrm{Re}(\tilde{\kappa}_3)\right]\phi_{3;(0,-)}^4\,,
\end{eqnarray}
with
\begin{eqnarray}
\nonumber&&\tilde{\kappa}_1\equiv\frac{1}{7}\kappa_0+\frac{4}{21\sqrt{5}}\kappa_2+\frac{6}{77}\kappa_4+\frac{100}{231\sqrt{13}}\kappa_6\,,\\
\nonumber&&\tilde{\kappa}_2\equiv\frac{1}{7}\kappa'_0+\frac{4}{21\sqrt{5}}\kappa'_2+\frac{6}{77}\kappa'_4+\frac{100}{231\sqrt{13}}\kappa'_6\,,\\
&&\tilde{\kappa}_3\equiv-\frac{1}{7}\kappa''_0-\frac{4}{21\sqrt{5}}\kappa''_2-\frac{6}{77}\kappa''_4-\frac{100}{231\sqrt{13}}\kappa''_6\,.
\end{eqnarray}
In the most general case, the dark matter is linear combination of $\phi_{3;(0,+)}$ and $\phi_{3;(0,-)}$, since the mass matrix for $\phi_{3;(0,+)}$ and $\phi_{3;(0,-)}$ shown in Eq.~(\ref{eq:massSQ_neutral}) is not diagonal, the dark matter self interactions can be easily extracted from Eq.~(\ref{eq:interactions_ab}). In the limit of both $M_B^2$ and $\lambda_3$ are real, the lightest one of $\phi_{3;(0,+)}$ and $\phi_{3;(0,-)}$ is the DM candidate, accordingly the self interaction can be read out straightforwardly.

For the electroweak scalar quintuplet $\phi_{2, 0}=(\phi_{2; (0, +)}+ i \phi_{2; (0,-)})/\sqrt{2}$, the the self interactions among the neutral fields $\phi_{2; (0, +)}$ and $\phi_{2; (0,-)}$ read as 
\begin{eqnarray}
\nonumber&&\frac{1}{4}\left[\tilde{\kappa}_1+2\mathrm{Re}(\tilde{\kappa}_2)+2\mathrm{Re}(\tilde{\kappa}_3)\right]\phi_{2;(0,+)}^4\\
&&-\left[2\mathrm{Im}(\tilde{\kappa}_2)+\mathrm{Im}(\tilde{\kappa}_3)\right]\phi_{2;(0,+)}^3\phi_{2;(0,-)}\\
\nonumber&&+\frac{1}{2}\left[\tilde{\kappa}_1-6\mathrm{Re}(\tilde{\kappa}_2)\right]\phi_{2;(0,+)}^2\phi_{2;(0,-)}^2\\
\nonumber&&+\left[2\mathrm{Im}(\tilde{\kappa}_2)-\mathrm{Im}(\tilde{\kappa}_3)\right]\phi_{2;(0,+)}\phi_{3;(0,-)}^3\\
\label{eq:interactions_ab}&&+\frac{1}{4}\left[\tilde{\kappa}_1+2\mathrm{Re}(\tilde{\kappa}_2)-2\mathrm{Re}(\tilde{\kappa}_3)\right]\phi_{2;(0,-)}^4\,,
\end{eqnarray}
with
\begin{eqnarray}
\nonumber&&\tilde{\kappa}_1\equiv\frac{1}{5}\kappa_0+\frac{2}{7\sqrt{5}}\kappa_2+\frac{6}{35}\kappa_4\,,\\
\nonumber&&\tilde{\kappa}_2\equiv\frac{1}{5}\kappa'_0+\frac{2}{7\sqrt{5}}\kappa'_2+\frac{6}{35}\kappa'_4\,,\\
&&\tilde{\kappa}_3\equiv\frac{1}{5}\kappa''_0+\frac{2}{7\sqrt{5}}\kappa''_2+\frac{6}{35}\kappa''_4\,.
\end{eqnarray}


\end{document}